\documentclass[twocolumn,showpacs,aps,prl]{revtex4}

\usepackage{graphicx}%
\usepackage{dcolumn}
\usepackage{amsmath}
\usepackage{latexsym}
\usepackage{morefloats}

\begin {document}

\title {Cyclic and Coherent States in Flocks with Topological Distance}
\author
{Biplab Bhattacherjee$^{1}$, K. Bhattacharya$^2$ and S. S. Manna$^{1}$}
\affiliation
{
\begin {tabular}{c}
$^1$Satyendra Nath Bose National Centre for Basic Sciences,
Block-JD, Sector-III, Salt Lake, Kolkata-700098, India \\
$^2$Department of Physics, Birla Institute of Technology and Science, Pilani - 333 031, Rajasthan, India
\end{tabular}
}
\begin{abstract}
      A simple model of the two dimensional collective motion of a group of mobile agents have been studied. 
   Like birds, these agents travel in open free space where each of them interacts with the first $n$ 
   neighbors determined by the topological distance with a free boundary condition. Using the same prescription 
   for interactions used in the Vicsek model with scalar noise it has been observed that the flock, 
   in absence of the noise, arrives at a number of interesting stationary states. One of the two most prominent 
   states is the `single sink state' where the entire flock travels along the same direction maintaining perfect 
   cohesion and coherence. The other state is the `cyclic state' where every individual agent executes a uniform 
   circular motion, and the correlation among the agents guarantees that the entire flock executes a pulsating 
   dynamics i.e., expands and contracts periodically between a minimum and a maximum size of the flock.
   We have studied another limiting situation when refreshing rate of the interaction zone is the fastest.
   In this case the entire flock gets fragmented into smaller clusters of different sizes.
   
      On introduction of scalar noise a crossover is observed when the agents cross over from a ballistic 
   motion to a diffusive motion. Expectedly the crossover time is dependent on the strength of the 
   noise $\eta$ and diverges as $\eta \to 0$.
   
      An even more simpler version of this model has been studied by suppressing the translational 
   degrees of freedom of the agents but retaining their angular motion. Here agents are the spins, 
   placed at the sites of a square lattice with periodic boundary condition. Every spin interacts with 
   its $n$ = 2, 3 or 4 nearest neighbors. In the stationary state the entire spin pattern moves as a whole
   when interactions are anisotropic with $n$ = 2 and 3; but it is completely frozen when the interaction
   is isotropic with $n=4$. These spin configurations have vortex-antivortex pairs whose density 
   increases as the noise $\eta$ increases and follows an excellent finite-size scaling analysis.
\end{abstract}

\maketitle
\vskip -0.5 cm 
\section {1. Introduction}

      Flights of a flock of birds or a swarm of bees are well known examples of living systems exhibiting 
   collective behavior which are often modeled by groups of self-propelled mobile agents \cite{reynolds,vicsek-lazar-rev,
   toner-tu,couzin}. These groups are called `cohesive' since each agent maintains a characteristic distance 
   from other agents and at the same time they are `coherent' since all agents move along a common direction. 
   In addition, the internal structure and organization of a flock is found to be complex and far from being 
   static. Examples include leader-follower relationships in bird flocks \cite{nagy-nature}, fission-fusion 
   events in fish schools \cite{croft} and social groups in human crowd \cite{helbing}. Such flocks often travel 
   over long duration covering distances much larger than the size of the flocks along arbitrary directions and therefore 
   effectively an infinite amount of space is available to the flock for it's motion. It is also known that while individual 
   agents often change their directions of flight the whole flock maintains motion along the same direction 
   in a stable fashion. 

\begin{figure*}
\begin{center}
\begin {tabular}{cc}
\includegraphics[width=6.0cm]{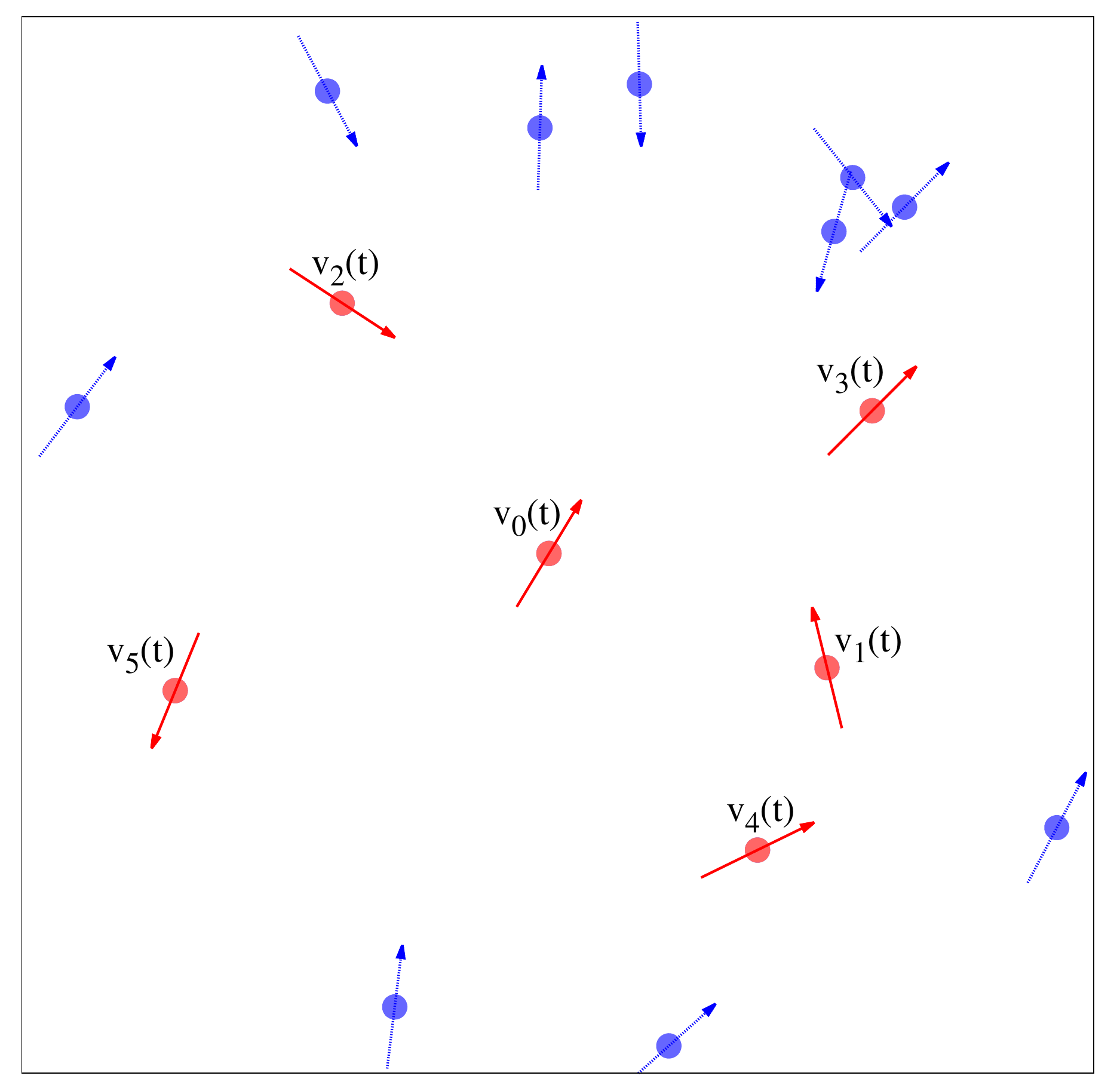} \hspace*{0.8cm}& \hspace*{0.8cm}
\includegraphics[width=6.0cm]{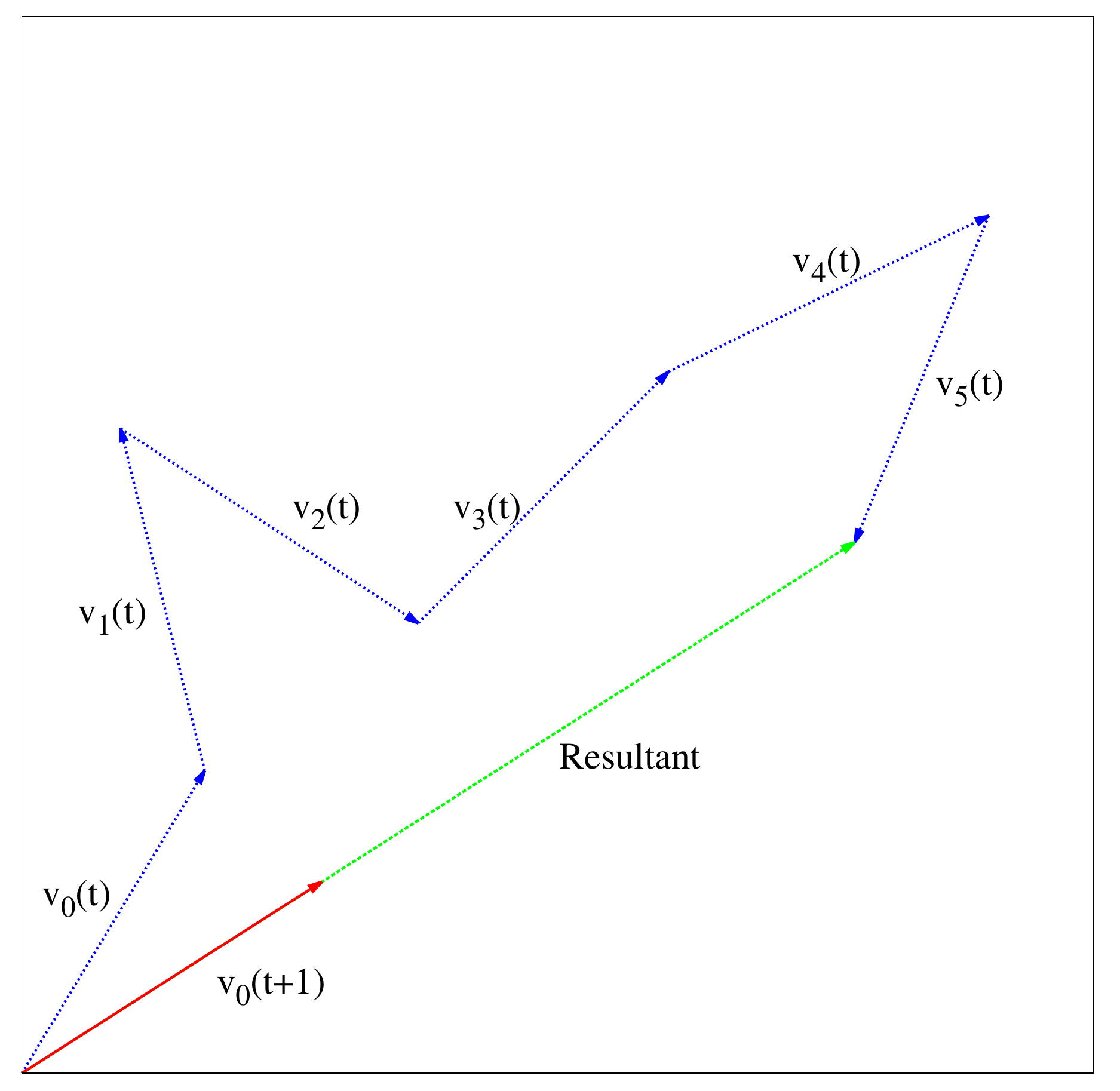} \\
{\bf (a)}  &  {\bf (b)}
\end {tabular}
\end{center}
\caption {(Color online) 
    (a) A flock of $N$ = 16 agents with $n=5$ neighbors in the interaction zone and at any arbitrary time $t$. 
    The central agent is denoted by the subscript 0 and the velocity vectors of this agent and its 5 neighbors in the 
    interaction zone are shown using red arrows.
    (b) In the next time step the velocity of agent 0 is calculated by Eqn. (1) which is along the resultant of
    all $n+1$ velocity vectors and has the magnitude $v$. 
}
\end{figure*}
   
      The generic feature of collective motion is, agents are short-sighted. While an individual agent's behavior 
   is influenced by the behavior of a small group of local agents around it, the whole group behaves in 
   unison. In other words a short range interaction among the agents may lead to a unique global behavior of 
   the entire group which implies the existence of a long range correlation among the agents. Therefore the 
   question is, given a random distribution of positions and velocities what kind of short range dynamics can 
   lead to global correlation reflected in cohesion and coherence among the agents.
  
      This question was first explored in an assembly of self-propelled particles, known as the Vicsek model 
   \cite {Vicsek}. Here particles (agents) are released at random locations within a unit square box on the 
   $x-y$ plane with periodic boundary condition and with random velocities. However, in the deterministic 
   motion the direction of velocity of each agent $i$ is oriented along the resultant velocity ${\bf v}_{Ri}$ 
   of all agents within an interaction zone (IZ) of range $R$ around $i$. In reality each agent may make an 
   error in judging the resultant direction of motion and this has been introduced in the stochastic version 
   of the model where noise is introduced by topping the orientational angle of ${\bf v}_{Ri}$ by a random 
   amount $\Delta \theta$. Each individual agent is then moved along the updated velocity direction. A 
   coherent phase is observed in the noise-free case with high agent densities. Moreover a continuous phase 
   transition is observed on increasing the strength of noise where the mean flock speed continuously 
   decreases to zero. However, facets like high density traveling bands occurring at low noise were revealed 
   in later studies \cite{Chate1,Chate2} and arguments were put in favor of a discontinuous transition.

   In a recent field study by the Starflag group on flocks of Starlings \cite {StarFlag}, it has been shown 
   that the interactions among the birds of a flock do not depend on the metric distance but on the topological 
   distance. More quantitatively they found that each bird interacts with a fixed number of neighbors, about six 
   or seven in number, rather than all neighbors within a fixed radial distance. Observing flocks of Starlings 
   the angular density distribution of neighboring birds have been found to be anisotropic e.g., a bird is more 
   likely to keep its nearest neighbor at its two sides rather than on the front and back. Fishes \cite{Gautrais} 
   have also been found to interact with neighbors determined by topological rules. Theoretical investigations 
   \cite{Chate3,Chate4} revealed that the behavior of topology based models are very different from metric based 
   models \cite{Heupe}. 
   
      The concept of graph theory based topology was, however, used \cite{Jad} to analyze the Vicsek model itself 
   from the  perspective of control theory. The metric distance based interactions were modeled using graphs with 
   ``switching topology''. Such studies also derived the conditions for the formation of coherent flocks for 
   agents with fixed topologies \cite{Tanner1}. The relevance of underlying graphs or networks on the nature of 
   collective motion has also been studied \cite{Bode1,Chate-nematics}.  

      The observations of the StarFlag group prompted us to study the collective motion of flocking phenomena in 
   two dimensions using the interactions depending on the topological distance. Most crucially we have obtained 
   very interesting stationary states which have not been observed before, mainly the cyclic states. At the same
   time an increasingly large number of states are found to be completely cohesive and coherent.
   The paper is organized as follows. In section 2 we describe our topological distance dependent model for 
   collective motion. The connectivity among such a collection of agents has been studied as the Random Geometric Graph in 
   section 3. The stationary states of such flocks have been studied in section 4, the two most prominent states 
   being the Single Sink State and the Cyclic State. The effect of the noise on the dynamics and the critical 
   point of transition have been studied in section 5. Study of the dynamics of the flock with the fastest refreshing 
   rate of the Interaction Zone has been done in section 6. A simpler version of the model with its vortex-antivortex 
   states have been studied on the square lattice in section 7. Finally we summarize and discuss in section 8. 

\begin{figure*}
\begin{center}
\begin {tabular}{cccc}
\includegraphics[width=4.0cm]{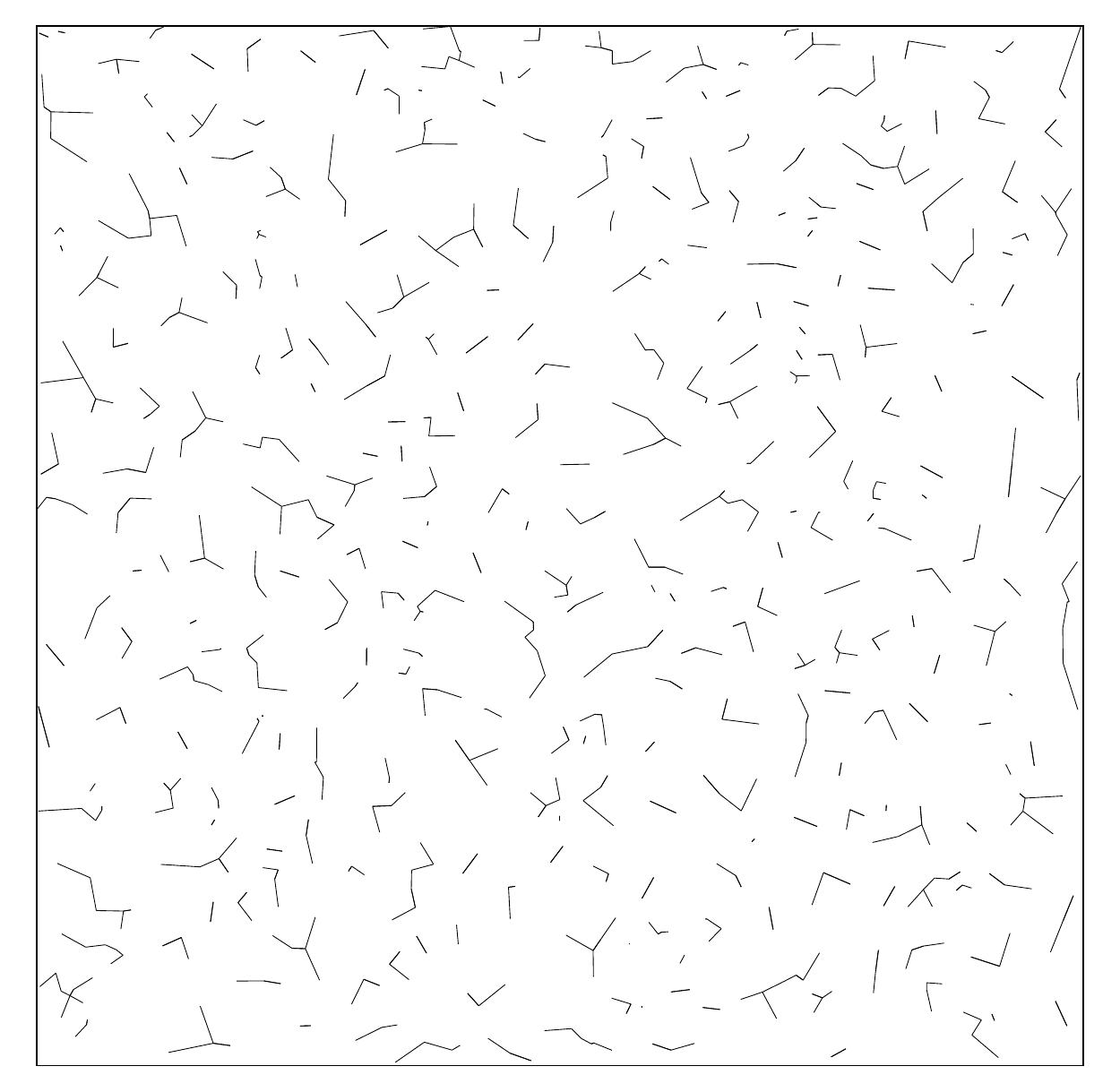} &
\includegraphics[width=4.0cm]{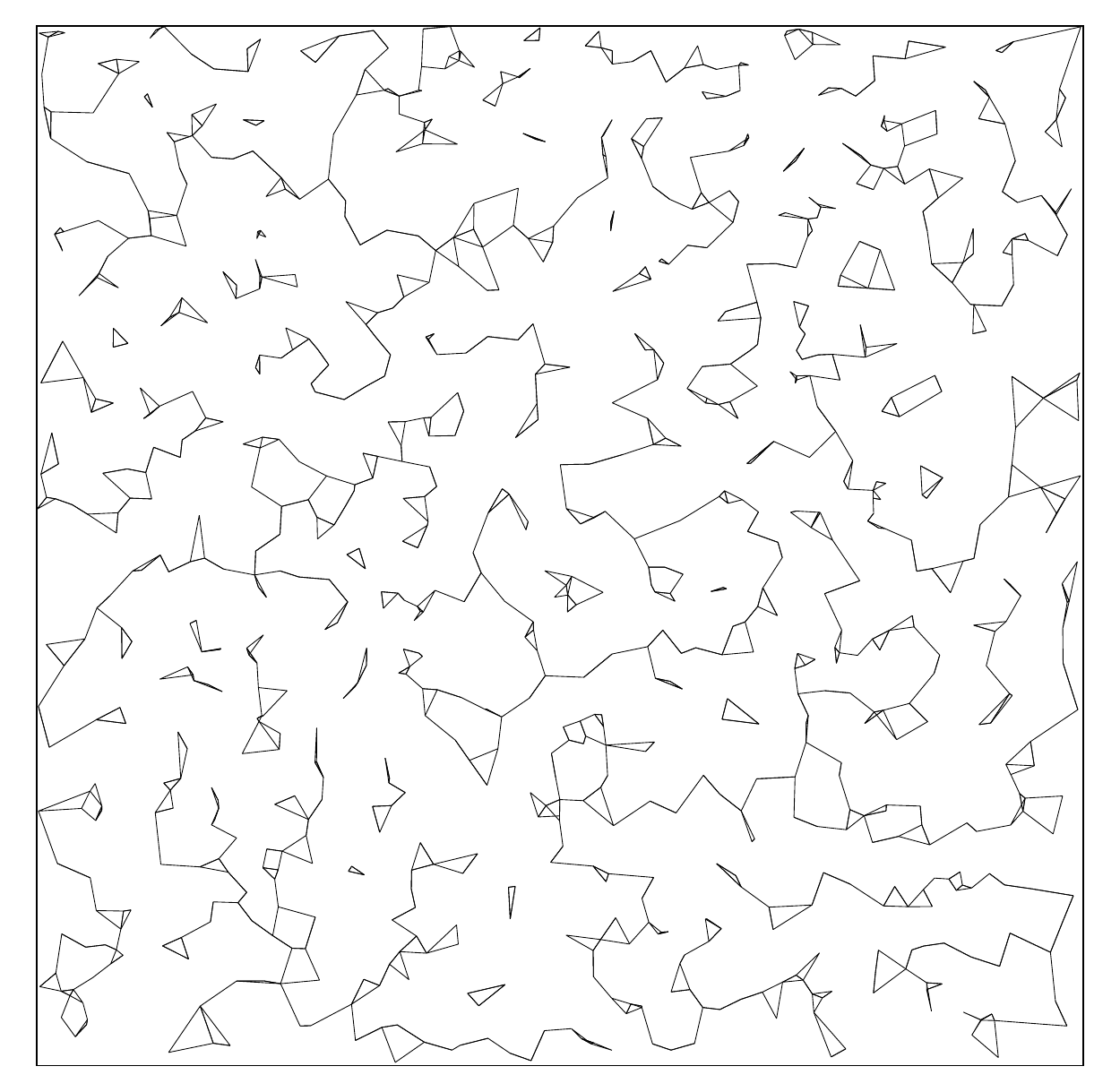} &
\includegraphics[width=4.0cm]{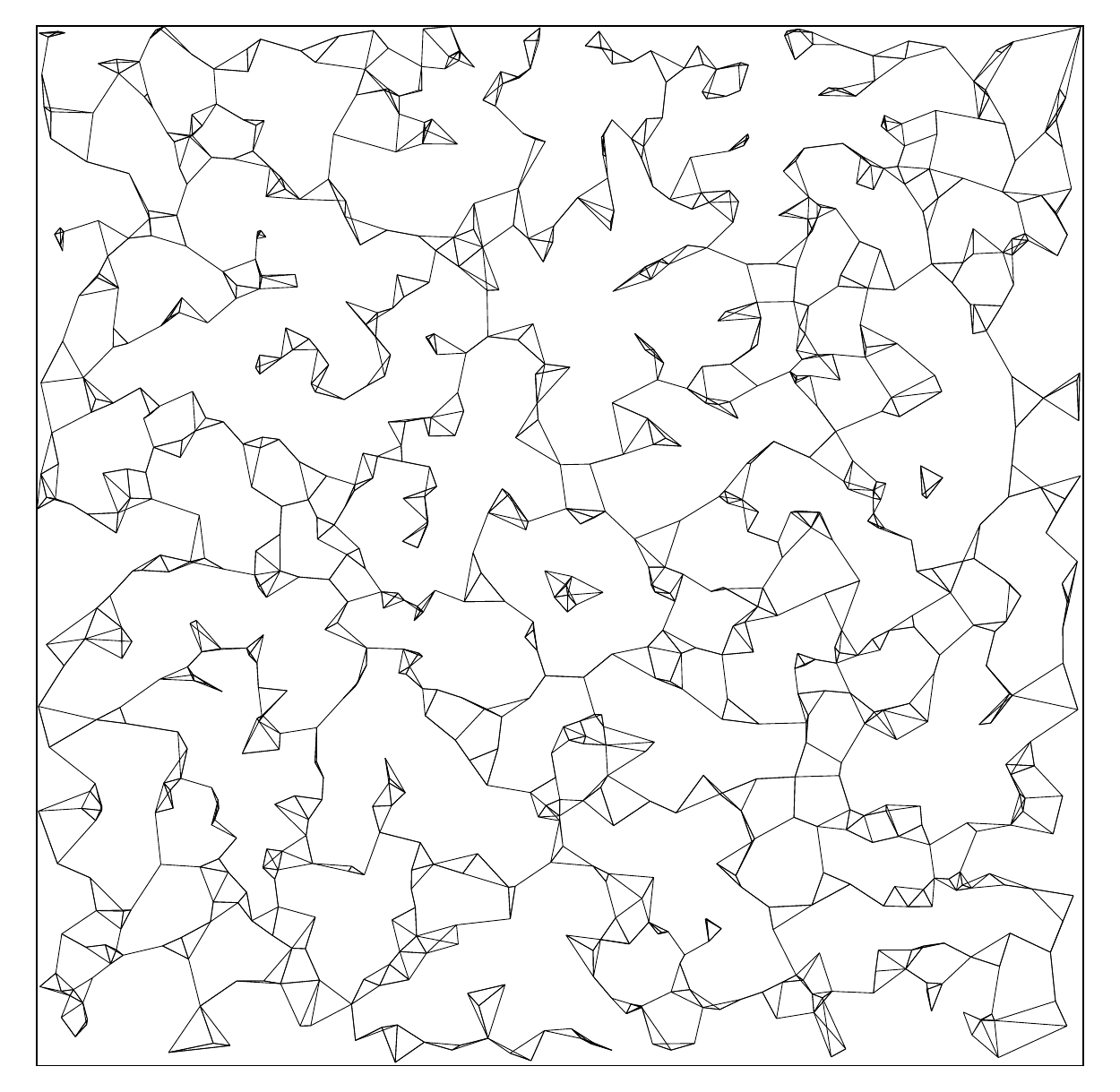} &
\includegraphics[width=4.0cm]{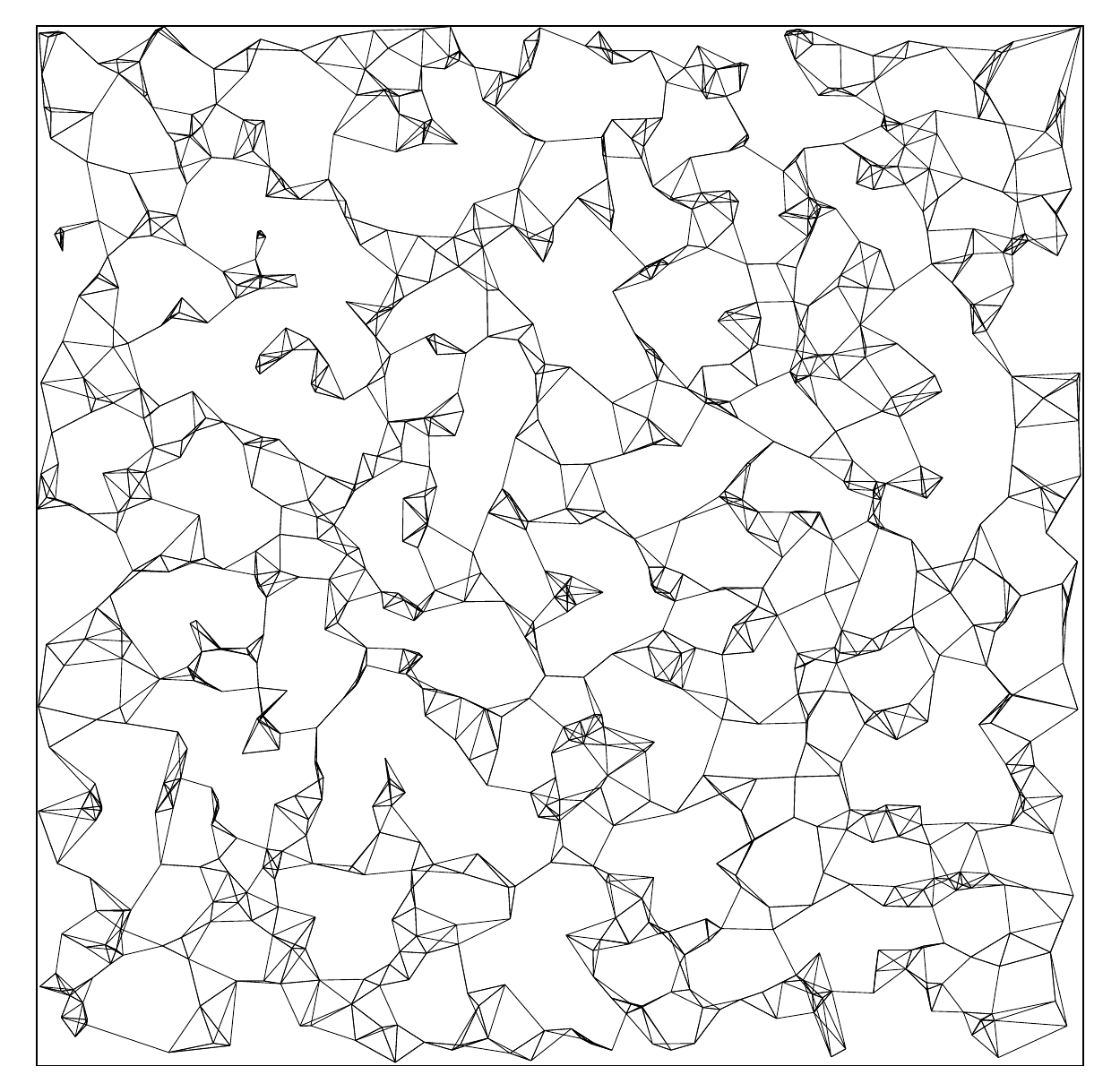} 
\end {tabular}
\end{center}
\caption{The undirected RGG with $N=1000$ nodes distributed randomly within a square box with a free 
boundary condition. Each node is linked to its $n$ nearest neighbors; $n$ = 1, 2, 3 and 4 increasing 
from left to the right. For small $n$ there are many components of the graph which merge with one 
another as $n$ increases. The largest component has sizes 9, 150, 988, 1000.}
\end{figure*}
\vskip -0.5 cm 
\section {2. Model}

      In our model the interaction zone has been defined in the following way. During the flight, each 
   agent $i$ interacts with a short list of $n$ other selected agents that constitute the IZ. It determines
   it's own velocity using the Eqn. (1) given below following a synchronous dynamics. In general, the agent 
   often refreshes the group of agents in 
   IZ. For example, at the early stage, when the flock is relaxing to arrive at the stationary state and also
   during some stationary states, the inter-agent distances change with time. Every time the IZ is refreshed, 
   we assume the criterion of selecting $n$ agents is that they are the first $n$ nearest neighbors of $i$. 
   We introduce at this point a ``refreshing rate'' which controls how frequently an agent updates it's IZ. 
   In this paper we study two limiting situations when these rates are slowest and fastest. In the slowest 
   rate, the agents do not change at all the the list of other $n$ agents in their IZs. The IZ for each 
   agent, constructed at the initial stage, remains the same ever after, even if $n$ initial neighbors of 
   an agent no longer remain nearest neighbors as time proceeds. The other limiting case is when the refreshing 
   rate is the fastest, the IZ is refreshed for every agent at each time step. The slowest case has been 
   discussed in sections 4 and 5. The fastest refreshing rates have been discussed in section 6. For the 
   spins on the square lattice discussed in section 7, these two cases actually mean the same since the 
   spins are firmly fixed at their lattice positions.

      The number $n$ of agents in IZ is considered as an integer parameter of the model. As in Vicsek model \cite {Vicsek} 
   the system is updated using a discrete time dynamics. While the speeds $v$ of all agents are always maintained to be the same, 
   the orientational angles $\theta_i$ of their velocities are updated by the direction of the resultant of 
   velocity vectors of all $n$ agents in the interaction zone and the agent $i$ itself (Fig. 1), 
\begin {equation}
   \theta_i(t+1)=\tan^{-1}[\Sigma_j \sin \theta_j(t) / \Sigma_j \cos \theta_j(t)]
\end {equation}
   where the summation index $j$ runs over all $(n+1)$ agents in IZ. The whole flock moves in the infinite space. 
   Following this dynamics, the flock reaches the stationary state after a certain period of relaxation time. It 
   is observed that the stationary state depends on the initial positions, initial velocities of the agents, as 
   well as the neighbor number $n$. Though a number of different stationary states have been observed, often the 
   state is a fixed point or a cycle. We have studied the statistical properties of these fixed points and cycles 
   and observed that cohesion and / or coherence are indeed present in different stationary states.
   
   There are two crucial differences of our model and the Vicsek's model which we summarize as follows:
   a) We have used the same prescription of interaction that had been claimed in the StarFlag 
   experiment. Accordingly, an agent here interacts with a fixed number $n$ of nearest neighbors 
   and not on all neighbors up to a fixed radial distance. It may be noted that, these two mechanisms are 
   essentially serving the same purpose. Both prescriptions keep the mutual interactions active among
   the local agents only. Therefore our choice  of neighbourhood and Vicsek model's range 
   of interaction are actually on the same footing.(b) No periodic boundary condition is imposed 
   in our model. Therefore agents move out in open space, yet they often form flocks that exhibit 
   considerable amount of cohesion and coherence. Compared to the Vicsek model, use of the free 
   boundary condition makes our model less restrictive.
   
      In the limiting case, when the IZ is refreshed at the slowest rate, it may appear that for
   an agent, any of the $n$ neighbours can be at an arbitrarily large distance. Certainly this 
   is not the case for a cohesive and coherent flock by definition. Moreover, we will see in the
   following that for most of the stationary states, the neighboring agents indeed remain within the
   close proximity of an agent.
   
\section {3. Random Geometric Graphs}

      At the initial stage $N$ agents are uniformly distributed at random locations within a unit square box on the 
   $x-y$ plane without periodic boundary condition. A random geometric graph (RGG) \cite {RGG} is constructed whose 
   vertices are the agents. At the same time for any arbitrary pair of vertices $i$ and $j$, $j$ is defined as a 
   neighbor of $i$ if it is among the $n$ vertices nearest to $i$. Then an edge is assumed to exist from $i$ to $j$. 
   This implies that the edges are in general `directed' since if $j$ is the neighbor of $i$ then $i$ may or may not 
   be the neighbor of $j$. Therefore the resulting graph is inherently a directed graph. However one can also define 
   a simplified version of the graph by ignoring the edge directions and consider the graph as an undirected graph. 
   In the following we refer such an undirected graph as the RGG.

\begin{figure}
\begin{center}
\includegraphics[width=7.0cm]{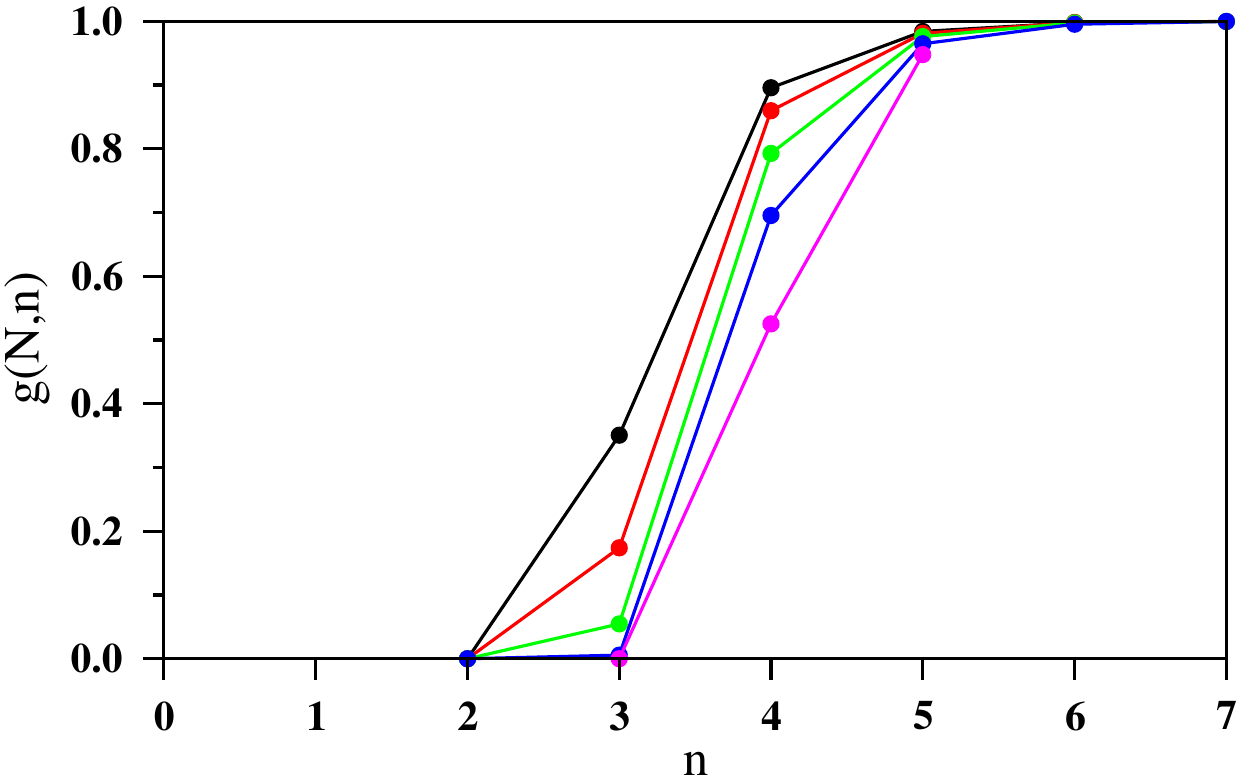}
\end{center}
\caption{(Color online)
The fractions $g(N,n)$ of single component connected graphs, in a sample of 1000 RGGs, are plotted against $n$. 
For $N$ agents this fraction grows as the number of neighbors $n$ is gradually increased. System sizes $N$ are 
256 (black), 512 (red), 1024 (green), 2048 (blue) and 4096 (magenta), increased from left to right. The number
of independent configurations used for each value of $n$ is 1000.
}
\end{figure}

       In Fig. 2 we exhibit the pictorial representation of an undirected RGG for $N=1000$ as $n$ is increased step by 
   step. For small values of $n$ the graph has many different components. As $n$ is increased the components grow 
   gradually in size, merge into one another and finally the RGG becomes a single component connected graph covering 
   all vertices for a certain value of $n$. Here we have shown four figures for $n$ =1, 2, 3 and 4. The randomly selected 
   positions of all vertices are exactly the same in these figures. The size of a component is measured by the number 
   of vertices in that component. In this figure the RGG becomes fully connected for $n$ = 4. 

      The structure and connectivity of RGG depend on the initial positions of $N$ vertices. Therefore we have first 
   studied how the fraction $g(N,n)$ of connected graphs grows with $n$ when the flock size $N$ is increased. For a 
   particular RGG the connectivity is checked using the `Burning Algorithm' \cite {Herrmann} where the fire, initiated 
   at an arbitrary vertex, propagates along the edges and finally burns all vertices if and only if the RGG is a single 
   component connected graph. 

      In Fig. 3 we show the plots of $g(N,n)$ against $n$ for different values of $N$. To find out if a minimum value
   of the neighbor number $n$ exists, one can artificially prepare a linear initial configuration of agents where each agent
   has it's right neighbor as the nearest one. This corresponds to $n=1$ but occurrence of such a configuration by random 
   selection of positions of the agents is extremely improbable. Numerically we find that for small $n$ the $g(N,n)$ takes
   vanishingly small values. However, on increasing $n$, $g(N,n)$ increases very rapidly and when $n$ is around 7, 
   $g(N,n) \approx 1$ i.e., nearly all configurations become connected. With increasing flock size $N$ the curves slowly 
   shifts to higher values of the neighbor number $n$.
   
      Only those flocks whose RGGs are single component connected graphs are considered for their dynamical evolution.
   The initial neighbor list is maintained for the entire dynamical evolution of the flock and is never updated even if
   all $n$ initial neighbors of an agent no longer remain nearest neighbors as time evolves. This means that the set of 
   agents' velocities $\{\vec v_i(t+1)\}$ is fully determined using the detailed knowledge of the set $\{\vec v_i(t)\}$. 
   Implication of this is, the positions and velocities are completely decoupled during the time evolution since the 
   actual positions of agents do not play any role to determine the velocities. Therefore the topological connectivity 
   of RGG remains invariant and is a constant of motion. 

\begin{figure*}
\begin{center}
\begin {tabular}{cc}
\includegraphics[width=7.0cm]{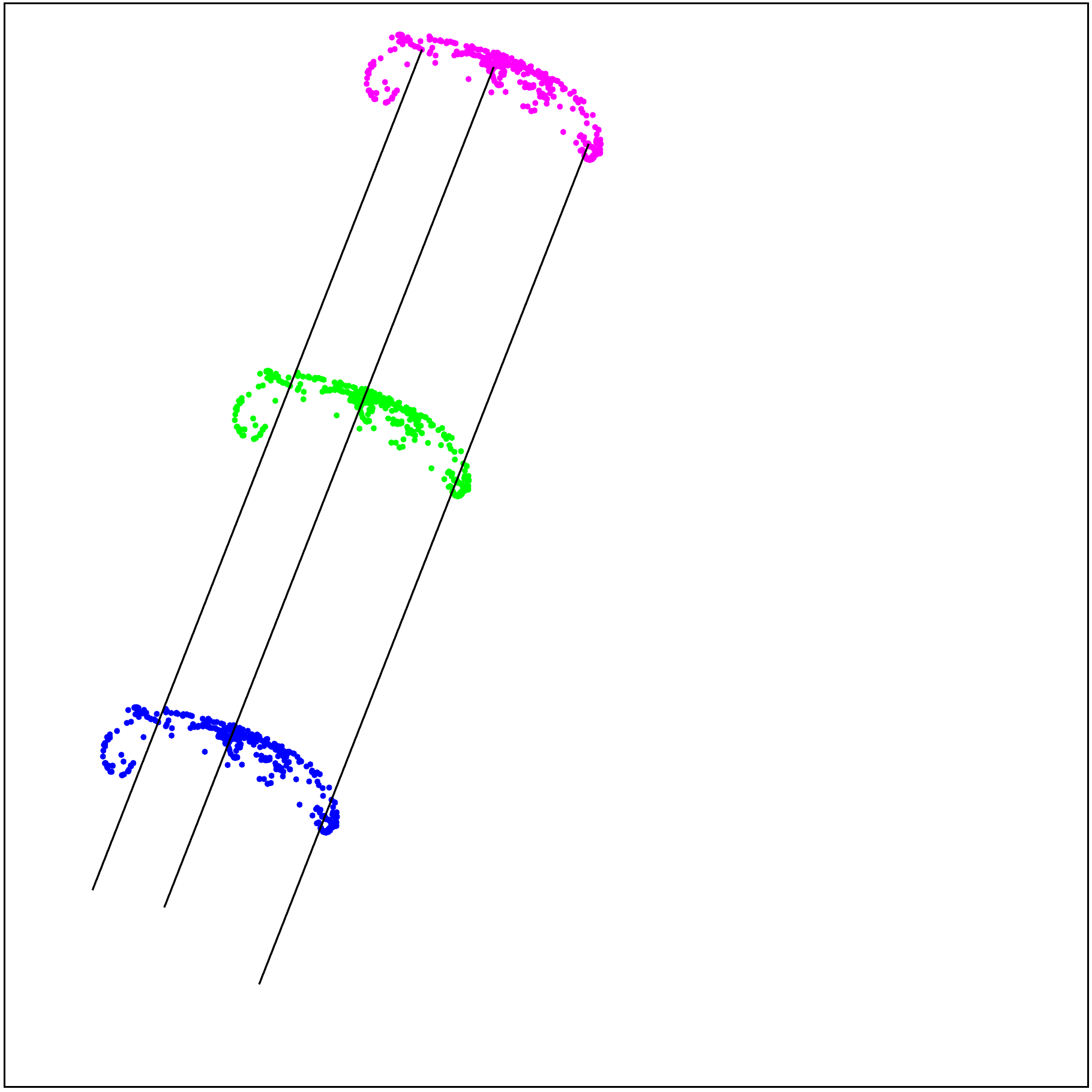} \hspace*{0.8cm}& \hspace*{0.8cm}
\includegraphics[width=7.0cm]{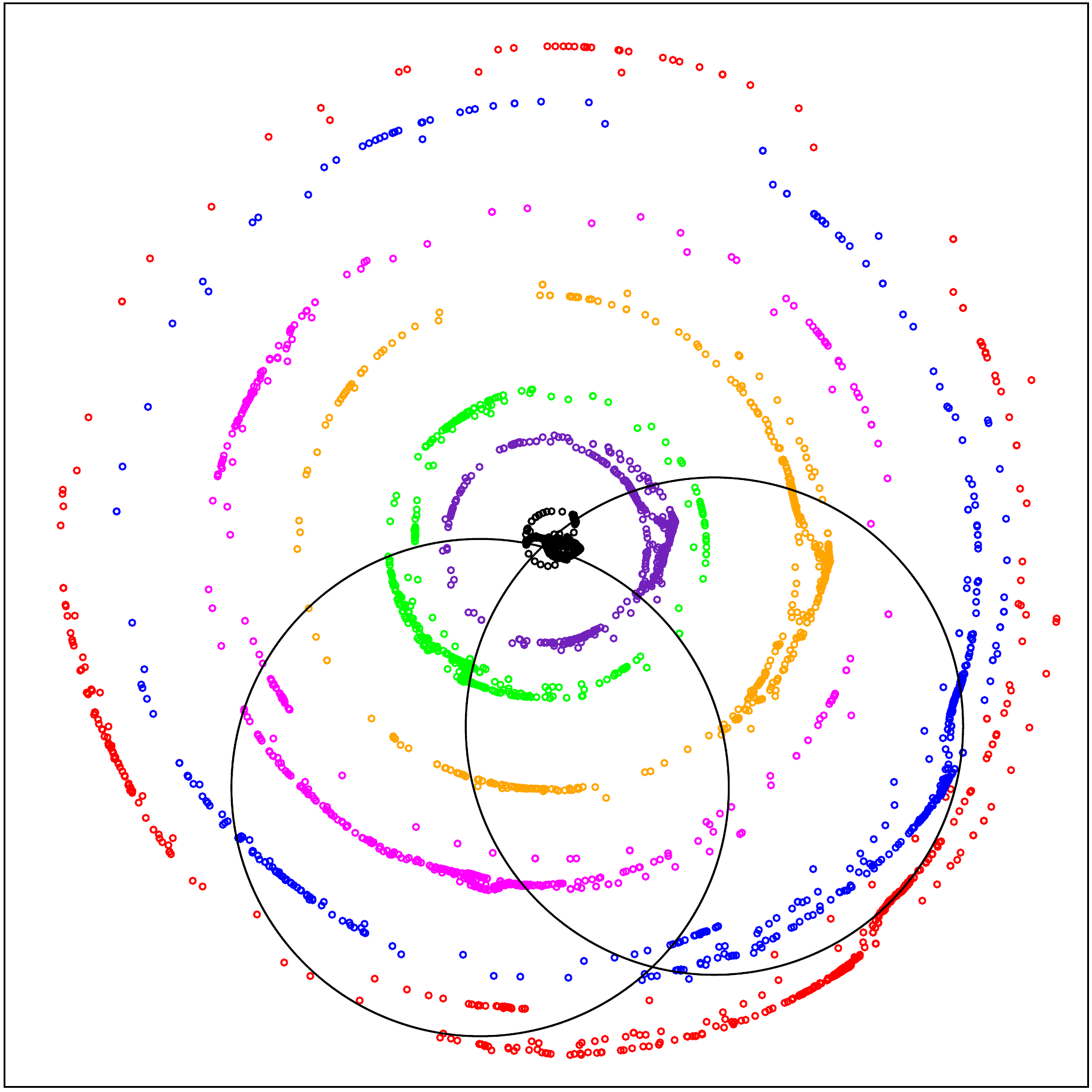} \\
{\bf (a)}  &  {\bf (b)}
\end {tabular}
\end{center}
\caption {(Color online) 
    Flocks of size $N$ = 512 and $n = 10$, moving with the speed of $v$ =0.03 without noise. 
(a) Single sink state: The fully cohesive and coherent motion of the flock is exhibited by 
    its position at three different instants: 10000 (blue), 11000 (green) and 12000 (magenta). 
    Three straight line trajectories of individual agents are also shown. The frame size is 
    $90 \times 90$ units. 
(b) Cyclic state: The stationary state pulsating flock has been shown at different time 
    instants: 173,000 (black), 176000 (red), 180000 (green), 188000 (blue), 192000 (brown), 
    197000 (violet) and 200000 (magenta). The time period is 28835. Two individual agents' 
    circular trajectories with radius $\approx 137.67$ are also shown.
    The frame size is $600 \times 600$ units.
}
\end{figure*}
   
\section {4. Stationary States}

\subsection {4.1. Single Sink States}

      Initially the $N$ agents are randomly distributed with uniform probabilities within the
   unit square box on the $x-y$ plane. If the corresponding RGG is fully connected then all agents are 
   assigned the same speed $v$ but along different directions. The angles $\theta_i$ of the velocity vectors 
   with respect to +$x$ axis are assigned by drawing them randomly from a uniform probability distribution between 0 and 
   $2\pi$. As time proceeds the agents soon come out of the initial unit 
   square box and spread out in the open two dimensional space. After some initial relaxation time the flock 
   arrives at the stationary state. One of the most common stationary state is the one where the flock is completely 
   coherent and cohesive. The entire flock moves along the same direction without changing the flock's spatial 
   cohesive shape and therefore $\theta_i(t) = {\it C}$ for all $i$ and are independent of time. 
   We call these states as the `Single Sink States' (SSS). Therefore this stationary state is a fixed point of the
   dynamical process. A picture of such a flock has been shown in Fig. 4(a). 

\begin{figure}
\begin{center}
\includegraphics[width=7.0cm]{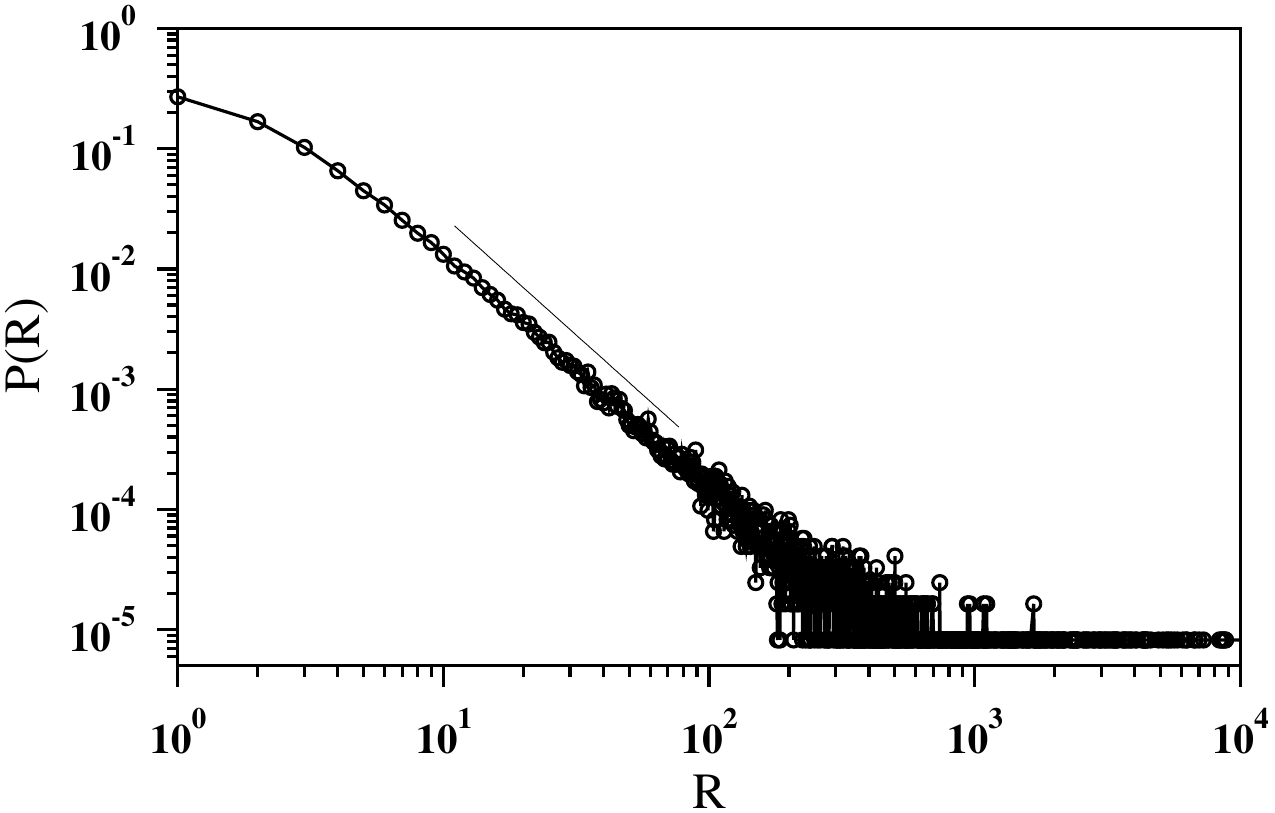}
\end{center}
\caption{
The probability distribution of the radii of the individual agent's circular orbits in the
cyclic states. The power law has an exponent of $\tau = 1.99(2)$. 
}
\end{figure}

\begin{figure*}
\begin{center}
\begin {tabular}{ccc}
\includegraphics[width=5.5cm]{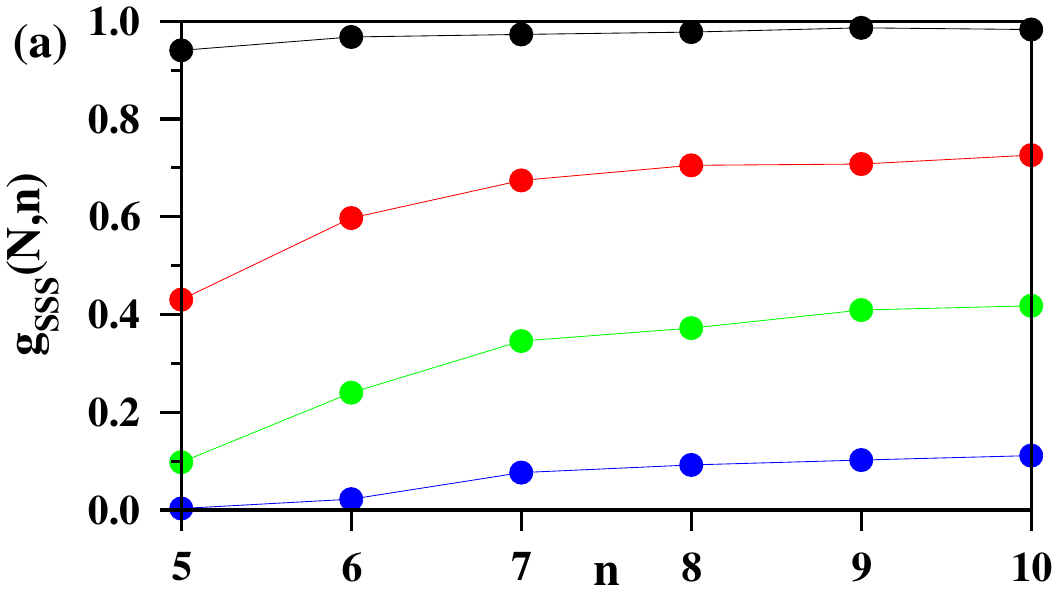} &
\includegraphics[width=5.5cm]{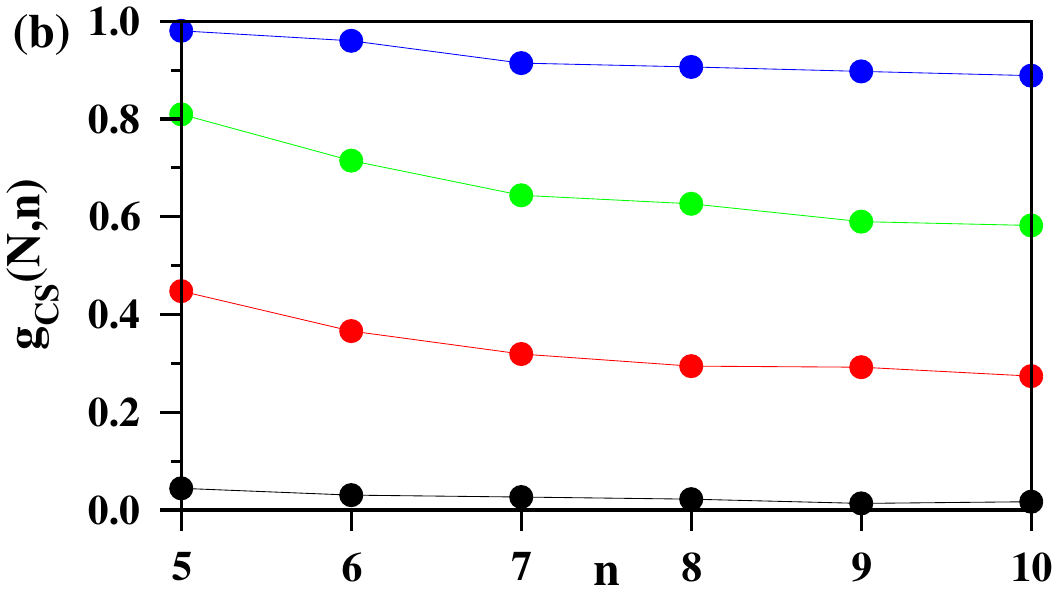} &
\includegraphics[width=5.5cm]{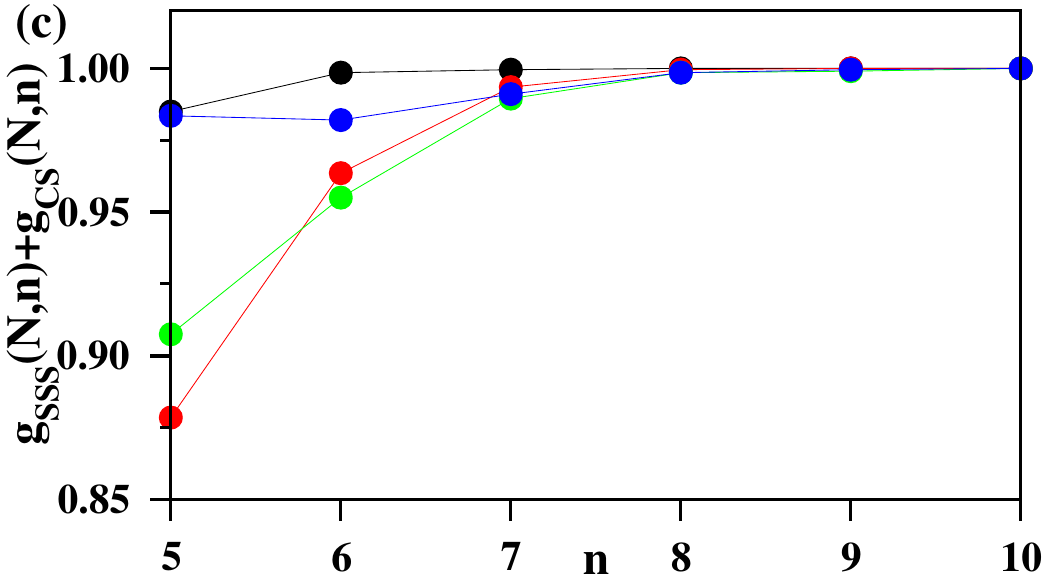} 
\end {tabular}
\end{center}
\caption{(Color online) The occurrence of two most prominent stationary states
when the neighbor number $n$ has been varied over a range from 5 to 10 and for
for different flock sizes $N$ = 64 (black), 256 (red), 512 
(green) and 1024 (blue). 
(a) The fraction $g_{SSS}(N,n)$ of SSS has been plotted against $n$.  
(b) The fraction $g_{CS}(N,n)$ of CS   has been plotted against $n$. 
(c) The sum of $g_{SSS}(N,n) + g_{CS}(N,n)$ has been plotted 
and it is seen that beyond $n \approx 8$ the sum is approximately unity.
}
\end{figure*}

\begin{figure*}
\begin{center}
\begin {tabular}{ccc}
\includegraphics[width=5.0cm]{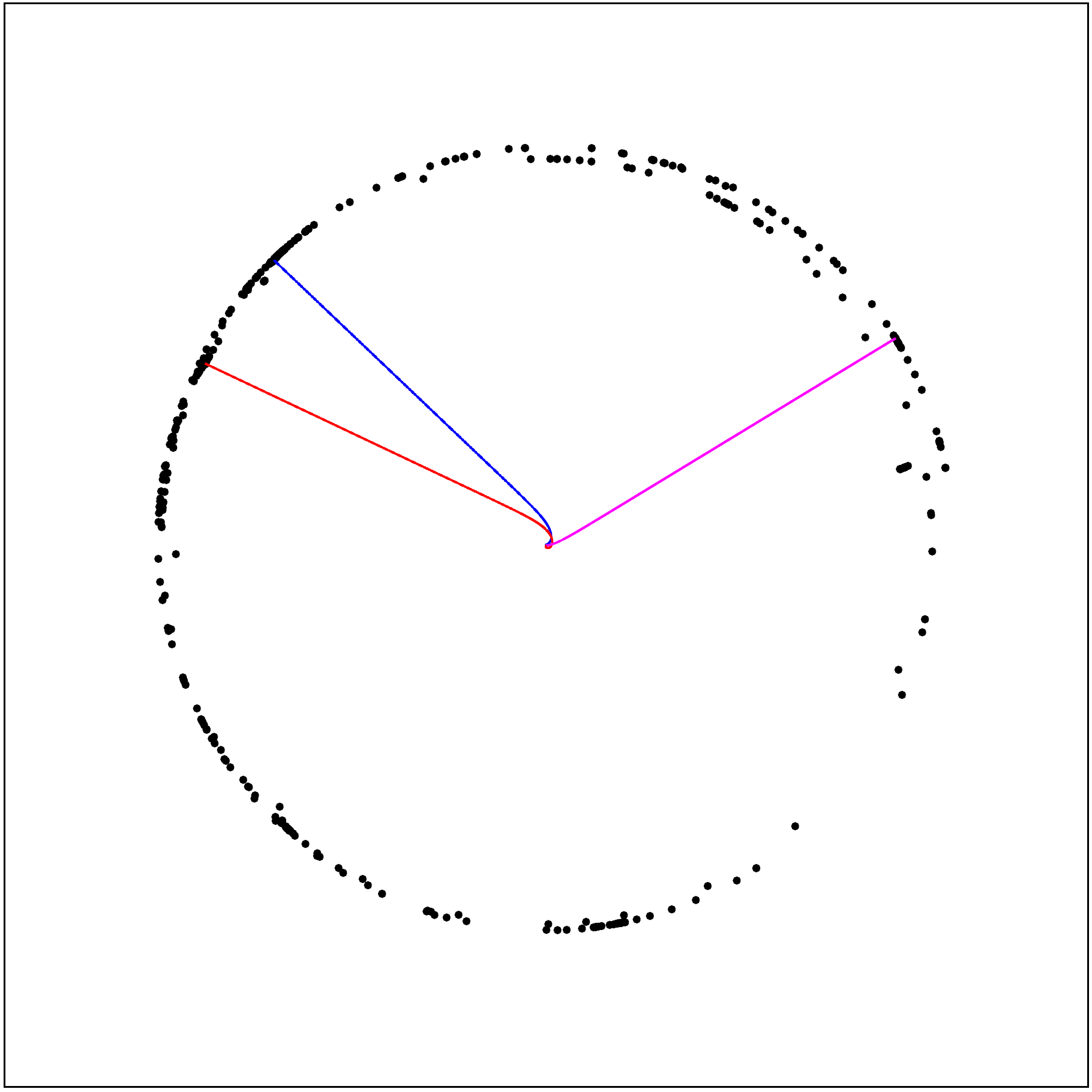} \hspace*{0.2cm}& \hspace*{0.2cm}
\includegraphics[width=5.0cm]{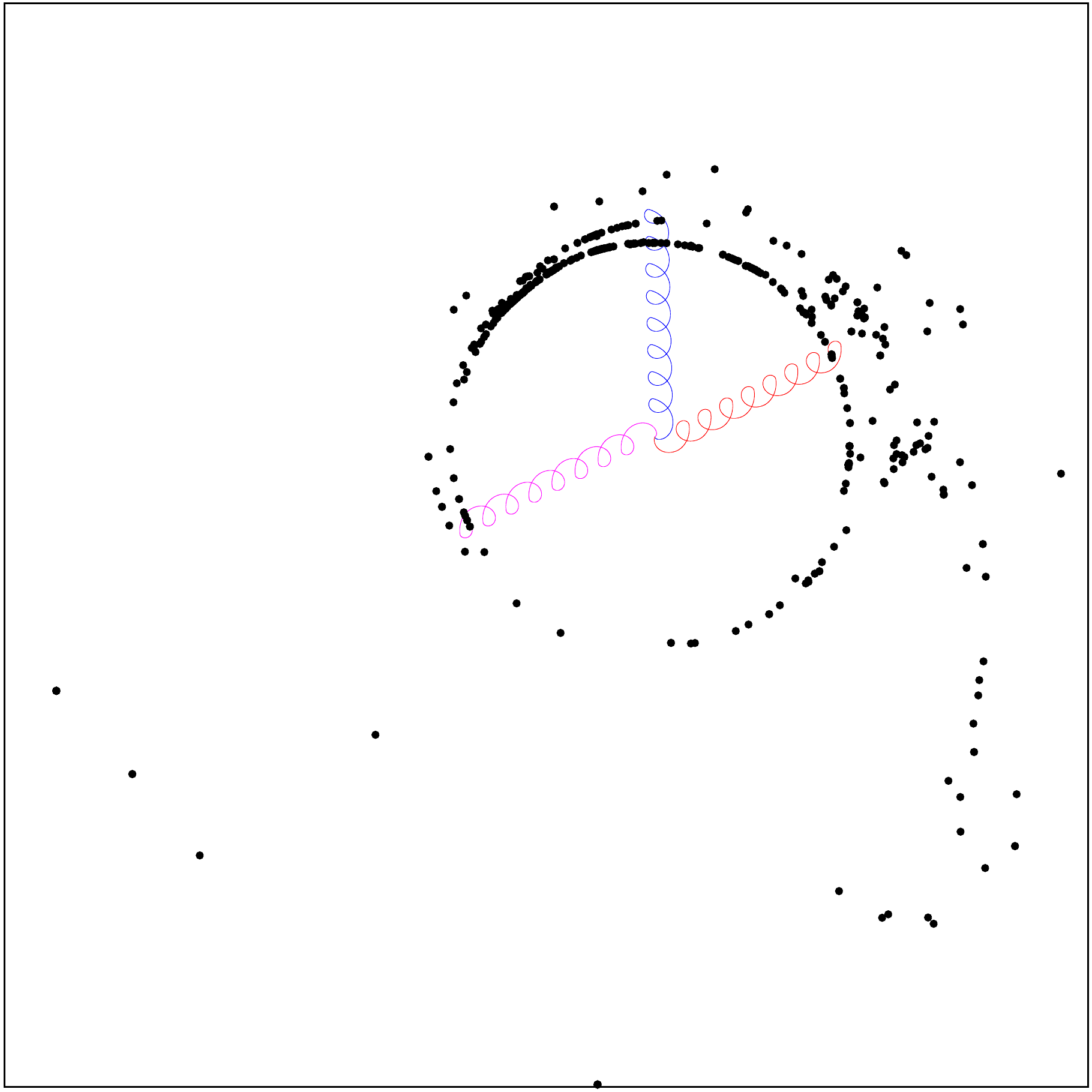} \hspace*{0.2cm}& \hspace*{0.2cm}
\includegraphics[width=5.0cm]{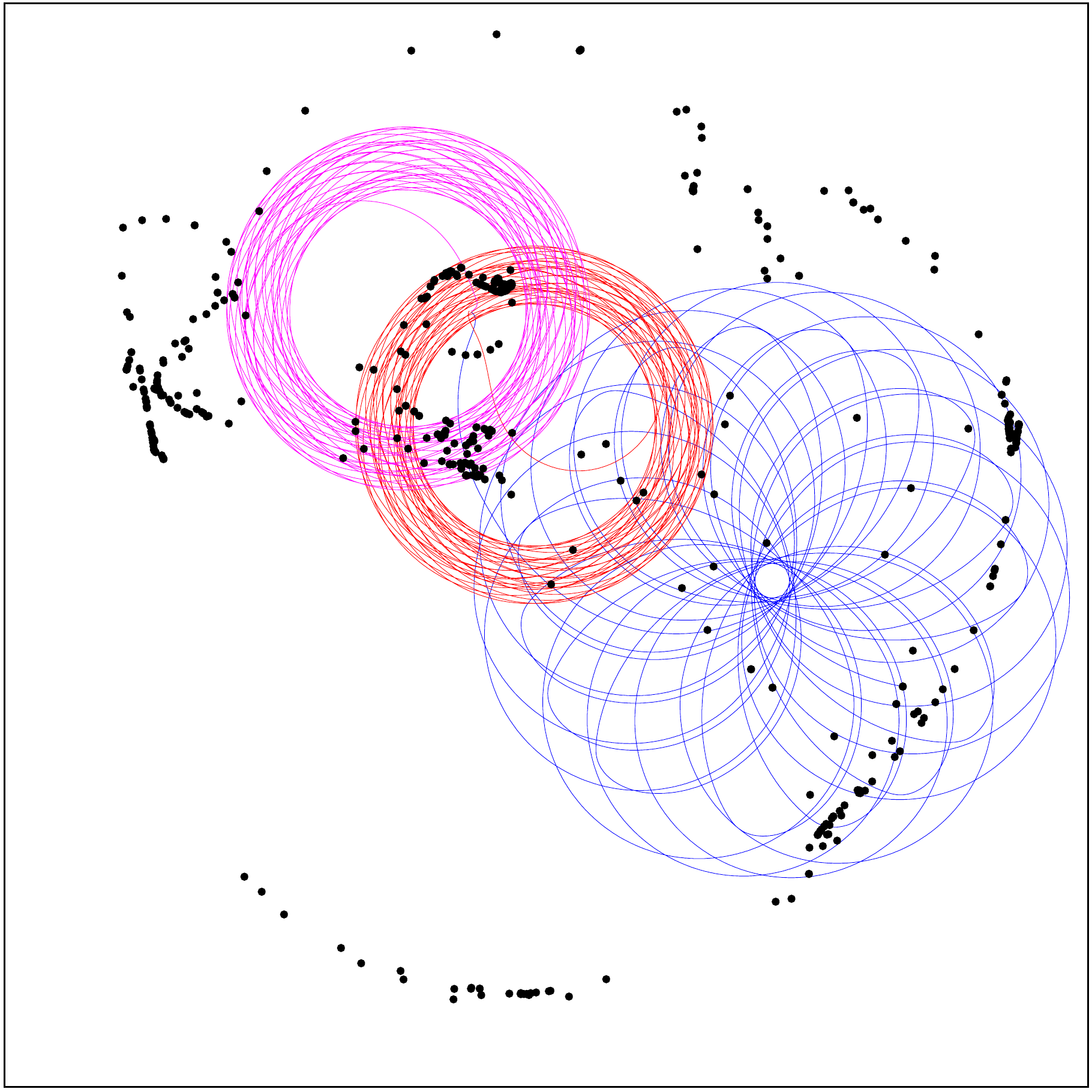} \\
{\bf (a)} \hspace*{0.2cm}& \hspace*{0.2cm} {\bf (b)} \hspace*{0.2cm}& \hspace*{0.2cm} {\bf (c)}
\end {tabular}
\end{center}
\caption {(Color online) Flocks of size $N$ = 512 and $n = 5$, moving with the speed of $v$ =0.03 without noise.
The positions of the agents are marked by black dots and three individual agent's trajectories 
are shown in each case by red, blue and magenta colors.
(a) Distributed sink state: Every agent moves along a fixed direction $\theta_i = {\it  C}_i$ of its own which is different in general 
from the directions of motion of other agents. 
(b) Cycloid state: In the stationary state the trajectory of each agent is a cycloid.
(c) Space-filling state: The trajectory of an agent never repeats itself but gradually fills up the
space between two concentric circles. The frame sizes are
40000, 15000 and 350 units respectively.} 
\end{figure*}

\subsection {4.2. Cyclic States}

      In cyclic states the velocity directions $\theta_i$ of all agents change at a constant rate in absence of 
   noise. Therefore the angular velocity $\dot \theta_i(t) = {\it D}$ for all agents is a constant of motion. 
   Each agent moves in a circular orbit of its own depending on its initial position, but their radii and time 
   periods are the same. Consequently the magnitude of the resultant velocity of the whole flock in the CS has 
   also a constant value but its direction changes with the same angular velocity. In addition typically the 
   shape of the flock is another circle though with some irregularities and interestingly its radius changes 
   periodically with the same period of individual agents. Therefore the whole circular flock pulsates, i.e., 
   periodically expands and contracts where each agent moves on its own fixed circular trajectory. We explain 
   this motion in Fig. 4(b) by plotting the flock at different instants of time and also show two individual 
   agent trajectories. We call these states as the `Cyclic States' (CS).

      For an arbitrary CS, let the probability that the radius of individual agent's circular trajectory between 
   $R$ and $R+dR$ be $P(R)dR$. Given that the uniform speed of the agents is $v$ and their angular velocities is 
   $\dot \theta$, the radius of the circular trajectory is $R=v/\dot \theta$. We have studied a large number of 
   such cyclic states and measured the radii of the agents' orbits. In Fig. 5 we show the probability distribution 
   of these radii which follows a power law distribution $P(R) \sim R^{-\tau}$ with $\tau = 1.99(2)$. 
   
      Throughout this paper we have used only one value of the agent speed, i.e., $v$ = 0.03. If the speed is reduced by a 
   certain factor a CS state remains CS but all the characteristic lengths are reduced by the same factor. The 
   radius of the circular orbit of every agent and also the size of the flock are reduced by the same factor,
   the time period remaining the same. Therefore it appears that even in the continuous limit of $v \to 0$, the 
   characteristic features of the flocks reported here remain same.
   
      Starting from the initial state, when the random positions and velocities are assigned to all agents, 
   the fractions of stationary states that exhibit the SSS and CS are estimated and are denoted by $g_{SSS}(N,n)$ 
   and $g_{CS}(N,n)$ respectively. In Fig. 6 these two quantities are plotted against the neighbor number $n$ for 
   different flock sizes $N$. For a certain $N$, $g_{SSS}(N,n)$ gradually increases with increasing $n$ (Fig. 6(a)). 
   For a given flock size $N$, however large, if the neighbor number $n$ is increased to $N-1$, then
   on using the dynamics mentioned in Eqn. (1) the stationary state flock must be both cohesive and perfectly coherent 
   i.e., $g_{SSS}(N,N-1) = 1$. No stationary state other than SSS can exist in this limiting situation. On the other
   hand when $n < N-1$ but $n$ is increased, then $g_{SSS}(N,n)$ also gradually increases and approaches the value of
   unity for any arbitrary value of $N$. At the same time, $g_{CS}(N,n)$ decreases with 
   $n$ for a fixed $N$ but increases with $N$ for a fixed $n$ (Fig. 6(b)). Finally in Fig. 6(c) we plot 
   the sum $g_{SSS}(N,n) + g_{CS}(N,n)$ which is less than unity for small $n$, but on increasing $n$, this 
   sum gradually increases and reached $\approx 1$ for $n = 8$ for all $N$. It is therefore concluded that 
   if the neighbor number is increased all other states gradually disappear and only SSS and CS states mostly dominate
   but ultimately for even larger value of $n$ it is the SSS state that only survives.
 
\subsection {4.3. Other states}

      In addition there are a number of other stationary states, few of them are 
   described below, but the list may not be exhaustive. 
   
   (i) Each agent has a constant velocity but their directions are different for 
   different agents. For example the $i$-th agent has its direction of velocity $\theta_i = {\it C}_i$. In this 
   case the agents, after some relaxation time, moves outward radially. The shape of
   the flock is approximately circular, again with some irregularities, and the radius of the
   flock increases at a uniform rate. We call these states as the
   `Distributed Sink States'(DSS). In Fig. 7(a) an example of the DSS has been shown.
   The position of the flock is shown at $t=$ 500000 and three agents' trajectories have been shown using different
   colors.
   
   (ii) In another type of stationary state the trajectories of the individual agents are very similar to cycloids.
   Each agent moves radially outward in a nearly cycloidal motion (Fig. 7(b)). A considerable number of agents form 
   a flock of circular shape but others are scattered around this circular flock. We call these as `Cycloid States'.
   
   (iii) Thirdly there can be rosette type stationary states. The trajectory of each agent is
   like a rosette which never closes and lies between two concentric circles. Consequently
   in the long time limit the trajectories fill the space between the two circles. This means that
   the mean separation between consecutive intersections of the agent trajectory with a radial section
   gradually vanishes as the trajectory evolves for a longer time. We call these as `Space-Filling States'. Three 
   such rosette trajectories and the position of the flock have been shown in Fig. 7(c).
   
   Few points may be mentioned here about the characteristics of the different stationary states. 
   For example a possible anisotropic effect on the stationary states may exist due to the choice of the unit square box
   for releasing the agents. Initially the
   positions of the agents are selected randomly within unit square box on the two dimensional plane.
   We have compared that if the agents' locations are selected randomly within a circle of radius 1/2,
   no appreciable change have been observed in the fractions of different stationary states.
   
   The center of mass of the entire flock has different kinds of trajectories in different stationary states.
   In SSS, the center of mass moves in a straight line exactly similar to all other agents.
   In CS, the trajectory of the center of mass is also a circle, but it's radius is not the same as the
   radius of the orbit of the individual agents, but it is some what larger.
   In DSS the dynamics of the center of mass is similar to that of the agents. Although, the shape of the 
   flock is approximately circular, but the fact that, at any instant the positioning of the agents on the 
   circumference is not uniform, makes the center of mass move radially outwards in a straight line. Motion is 
   indeed unbounded. In cycloid states, the trajectory of the center of mass is also a cycloid and radially 
   outwards. The motion here is also unbounded. In space filling states, the trajectory of the center of mass 
   is rosette type. In this case the trajectory is bounded.
   
   How sensitive are the final stationary states on the choice of the random initial values of $\{x_i,y_i\}$ 
   and $\{\theta_i\}$ for the $N$ agents? To study this point, we tried with a flock that evolves from a certain
   initial configuration that evolves to a CS. Now we again evolve the same flock, but this time we slightly 
   change the initial configuration randomly by $x_i=x_i+a.10^{-4}.r$ and $y_i=y_i+a.10^{-4}.r$ where $r$ is a
   random number. The directions $\{\theta_i\}$ of the velocity vectors are maintained the same. We then tune 
   $a$, and found that with $0 < a < 1.70$ the stationary state is still a CS, but with different values of
   the orbit radius. When $a \ge 1.75$ the stationary state becomes SSS. We conclude that with some amount
   of perturbation the character of the stationary state remains same, but with even stronger perturbation
   the stationary state changes.
   
   A preliminary calculation with our model in three dimensions shows the following features.
   In general, to obtain connected graphs, the value of $n$ needs to be large compared to what is required 
   in two dimensions. Almost always the dynamics leads to a SSS in the steady state. We did not find any 
   other state starting from random initial conditions.

\section {5. Dynamics in presence of noise}

      Studying the role of noise on the dynamics of the flock is very crucial. It is
   assumed that every agent makes a certain amount of error in judging the angle
   of its velocity vector at each time step. More precisely given the angles $\theta(t)$ 
   of velocity vectors of all $n+1$ agents within the interaction zone at time $t$, it first
   calculates the resultant of these vectors using the Eqn. (1). It then tops up this angle
   by a random amount $\zeta(\eta)$ which is uniformly distributed within $\{-\eta/2,\eta/2\}$.
   Therefore the modified Eqn. (1) reads as:
\begin {equation}
   \theta_i(t+1)=\tan^{-1}[\Sigma_j \sin \theta_j(t) / \Sigma_j \cos \theta_j(t)] + \zeta(\eta).
\end {equation}
   The role of the noise is to randomize the deterministic dynamics and quite expectedly the stationary state
   structures of the flocks exhibited in the SSS and CS patterns are gradually
   lost. We have studied the effect of noise on both these states by gradually increasing the
   strength of the noise $\eta$. In both cases we use a flock of $N$ = 512 agents, each of them
   interacts with $n=10$ nearest neighbors and travel with speed $v=0.03$. Initially all of them
   are released within the square box of size unity. We first run the dynamics without
   any noise i.e., $\eta=0$ and ensure that the stationary state pattern is indeed a single
   sink state. As the dynamics proceeds we calculate the maximal distance $R_m(t)$ and the average distance $R_a(t)$ of an 
   agent from the center of mass $(x_c(t),y_c(t))$ of the flock. In Fig. 8(a) we
   plot these two quantities against time and they are exactly horizontal curves which are
   the signatures of the SSS state. These simulations are then repeated for $\eta > 0$ and
   the variations of $R_m(t)$ and $R_a(t)$ have been shown in Figs. 8(b), 8(c) 
   and 8(d) for $\eta = 0.2, 0.5$ and 1 respectively. In all four cases
   the flock starts with the same positions and velocities of the agents. It is seen that on increasing the strength
   of noise the stochasticity gradually sets in and the variations of $R_m(t)$ and $R_a(t)$ 
   gradually become random. A similar plot has been exhibited in Fig. 9 for the CS
   but for only a single flock. The Fig. 9(a) shows the zero noise case and the curves are
   periodic. However when the noise level is increased (Figs. 9(b)-(c)) it distorts the periodicity. With small
   values of $\eta$ the variations are slightly distorted from the periodic variations, however
   with larger strength of noise the distortion is much more. Finally for $\eta=1$ the fluctuations
   look random and similar to those in the SSS (Fig. 9(d)).
   
      Next we calculated the mean square displacement $\langle r^2(t,\eta) \rangle$ from the origin as time
   passes. The averaging has been done for a single agent within a flock and over many such independent 
   flock samples. The noise strength has been varied over a wide range of values. In Fig. 10 we have displayed 
   the variation of $\langle r^2(t,\eta) \rangle$
   against the time $t$ using a $\log - \log$ scale for six different values of $\eta$. Here again we 
   consider flocks with fully connected RGGs. On the other hand with zero noise these configurations may lead to any of the 
   possible stationary states. Simulating up to a maximal time of $T=10^8$ we observed a cross-over
   behavior in the mean square displacement. When $\eta$ is very small $\langle r^2(t,\eta) \rangle \sim t^{2\kappa}$
   with $\kappa \approx 1$ which implies that the flock maintains a ballistic motion at the early stage, i.e., 
   the coherence is still maintained during this period. On the other hand after a long time 
   one gets $\kappa = 1/2$ which indicates the diffusive behavior. This implies that even if a little noise is applied
   for a long time, the effect of the noise becomes so strong that the flock can no longer maintain a cohesive and
   coherent structure any more and agents diffuse away in space. Therefore for any  
   value of $\eta$ there is a cross-over from the ballistic to diffusive behavior. Consequently
   a cross-over time $t_c(\eta)$ can be defined such that for short times $t << t_c(\eta)$ the dynamics
   is ballistic with $\kappa=1$ and for $t >> t_c(\eta)$ the dynamics is diffusive with $\kappa=1/2$. In Fig. 10 we show
   this behavior and observe that the crossover time depends explicitly on the value of $\eta$ and diverges as
   $\eta \to 0$. The value of $t_c(\eta)$ has been estimated by the time coordinate of the
   point of intersection of fitted straight lines in the two regimes of Fig. 10: for $t << t_c(\eta)$ and for $t >> t_c(\eta)$.
   The value of $t_c(\eta)$ so estimated diverges as $t_c(\eta) \sim \eta^{-2.52}$.
   
      This transition is more explicitly demonstrated using a plot of the Order Parameter (OP) $M(\eta)$
   against $\eta$ (Fig. 11). The OP is defined as the time averaged magnitude of the resultant of all agents'
   velocity vectors, scaled by its maximum value
\begin {equation}
M(\eta) = \langle |\Sigma^N_{j=1} {\bf v}_j| \rangle/(Nv)
\end {equation}
   where $\langle ... \rangle$ denotes the time average over a long period of time in the stationary state.
   In Fig. 11 (a) we have plotted $M(\eta)$ against $\eta$ at an interval of $\Delta \eta = 0.1$ for different 
   system sizes from $N$ = 64 to
   1024. In these simulations the initial conditions are chosen to be completely coherent so 
   that the velocity vectors of all agents are in the same direction. In the absence of noise this
   situation is maintained and $M(0) = 1$ at all times for all system sizes. However for $\eta > 0$ noise
   sets in, but in the stationary state one still gets a non-zero OP. On further increasing 
   $\eta$ the OP decreases monotonically and ultimately vanishes. Therefore there exists a critical
   value $\eta_c(N)$ of the noise parameter where the transition from the ordered state
   to disordered state takes place. It is observed in Fig. 11(a) that as the system size $N$ is enlarged
   the transition becomes more and more sharp and shifts to the regime of small $\eta$.
   Further we have calculated the Binder cumulant $G(\eta)=1-\langle M^4(\eta) \rangle/3\langle M^2(\eta) \rangle^2$ 
   and plotted against $\eta$ in Fig. 11(b) for the same system sizes \cite {Binder}. The value of $G(\eta)$ drops from a constant
   value of around 2/3 at the small $\eta$ regime to about $1/3$ for large values of $\eta$.

\begin{figure*}
\begin{center}
\begin {tabular}{cc}
\includegraphics[width=7.5cm]{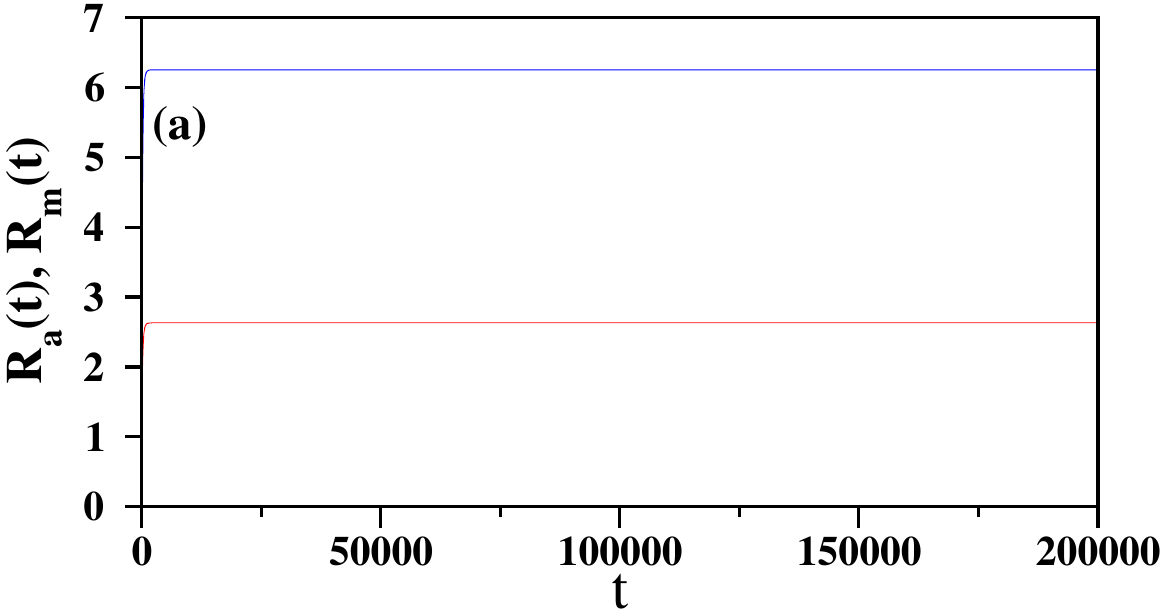} \hspace*{0.4 cm} & \includegraphics[width=7.5cm]{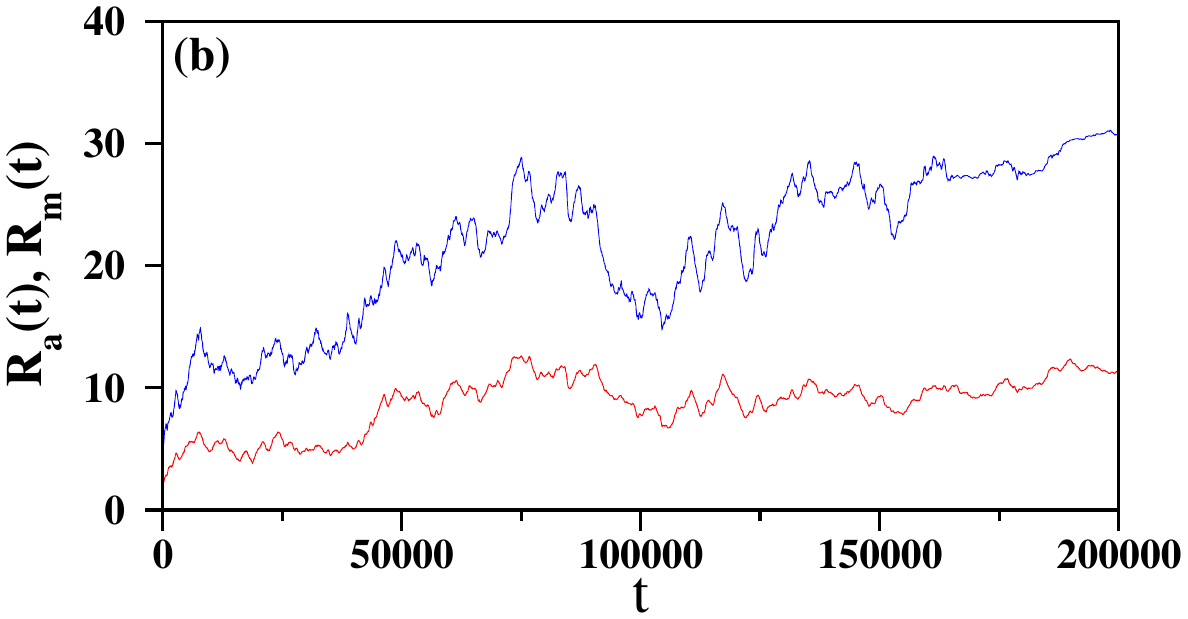} \\
\includegraphics[width=7.5cm]{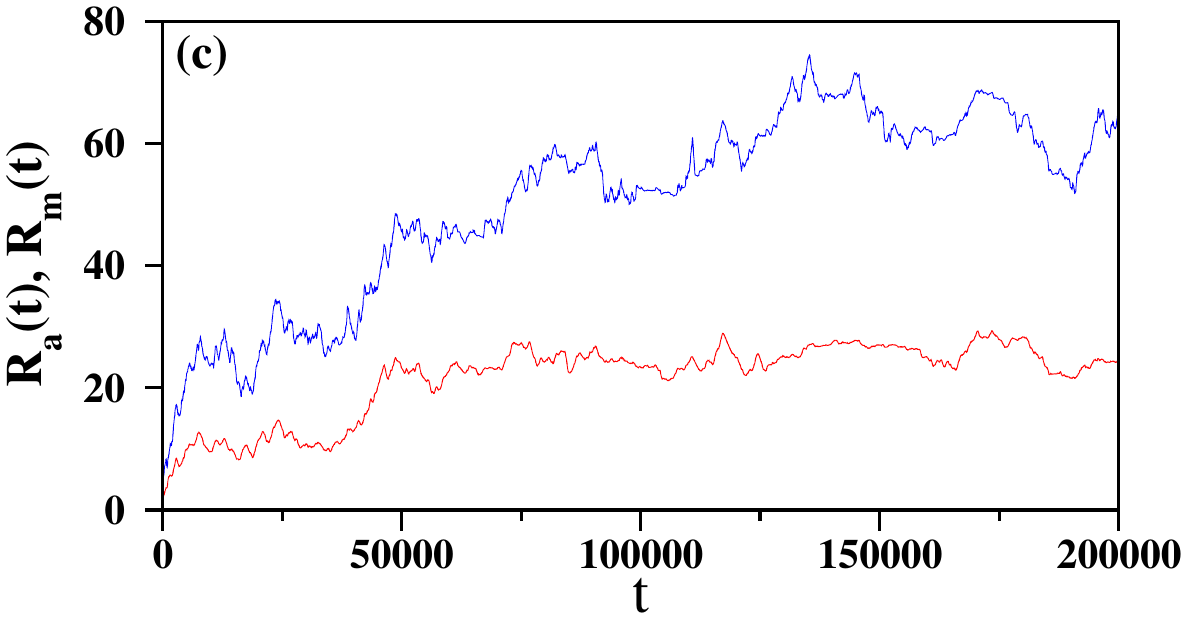} \hspace*{0.4 cm} & \includegraphics[width=7.5cm]{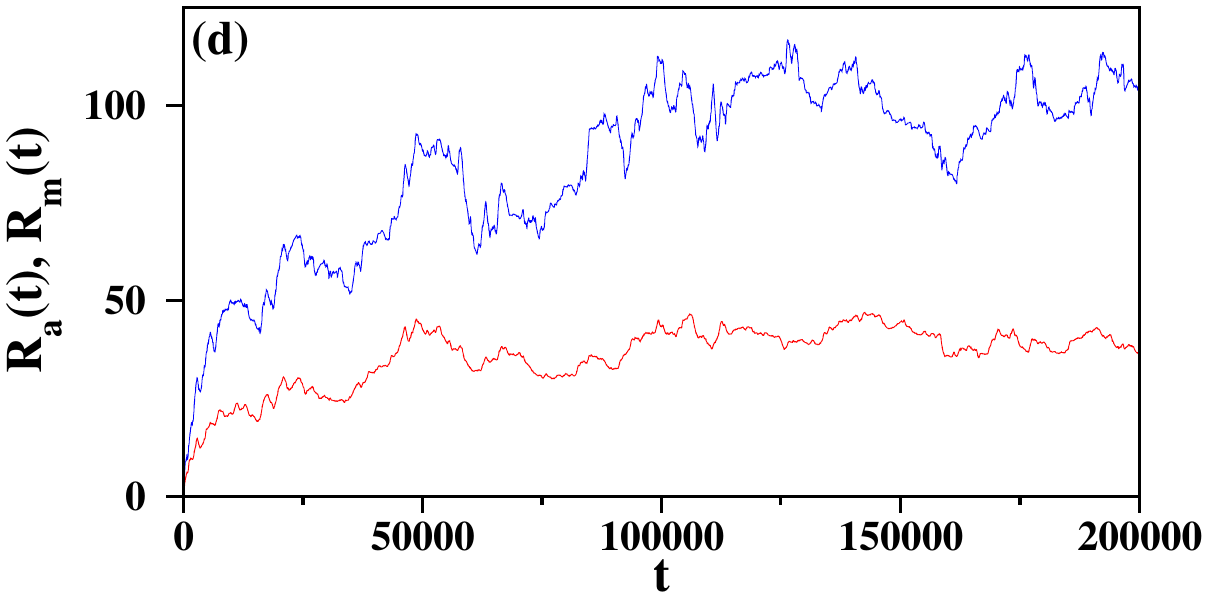} \\
\end {tabular}
\end{center}
\caption{(Color online)
Effect of noise is exhibited on a single flock ($N$ = 512 agents, each having $n$ = 10 neighbors) which 
goes to a single sink state without any noise. Variation of the maximal radius $R_{m}$ (blue) and the average 
radius $R_{a}$ (red) have been shown with different strengths of the noise parameter: (a) $\eta = 0$, 
(b) 0.2, (c) 0.5 and (d) 1. The initial positions and velocities are same in all four cases.
}
\end{figure*}
   
      The transition point $\eta_c$ can be estimated in the following two ways. For each curve in Fig. 11(a)
   we calculate the value of $\eta_{1/2}(N)$ for which $M(\eta)=1/2$. We define $\eta_{1/2}(N)$ is the
   characteristic noise level where the transition takes place. By interpolation of the plots in Fig. 
   11(a) of the points around $M(\eta)=1/2$ we have estimated $\eta_{1/2}(N)$. These estimates are
   then extrapolated in Fig. 11(c) as:
\begin {equation}
\eta_{1/2}(N) = \eta_{1/2}(\infty) + AN^{-1/\nu}.
\end {equation}
   On tuning the trial values of $\nu$ very slowly we found that for $\nu \approx 2.86$ the error in the
   least square fit of the above finite-size correction formula is minimum. Therefore the extrapolated 
   $\eta_{1/2}(\infty) \approx 1.70$ is the critical noise strength $\eta_c$ according to our estimate.
   A similar calculation has also been done using the Binder cumulant. From this calculation we estimated
   $\eta_c=1.82$ and $\nu = 3.50$. The differences between the two estimates are considered as the error
   in the measured values which are 0.12 and 0.64 for $\eta_c$ and $\nu$ respectively.   
   
\begin{figure*}
\begin{center}
\begin {tabular}{cc}
\includegraphics[width=7.5cm]{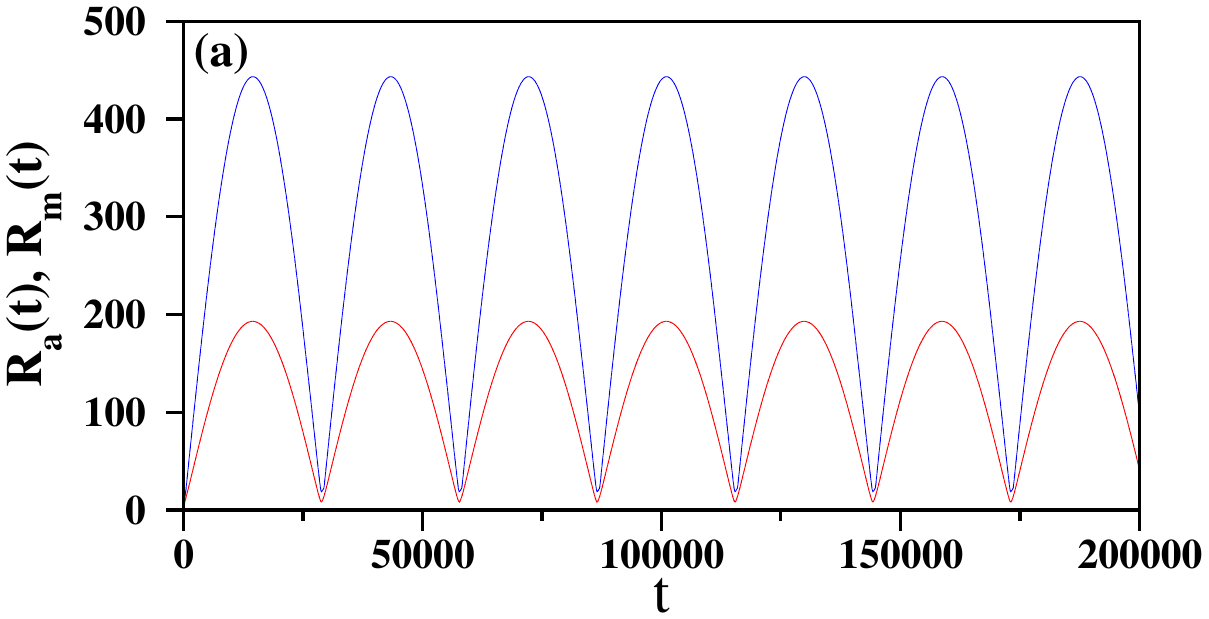} \hspace*{0.4 cm} & \includegraphics[width=7.5cm]{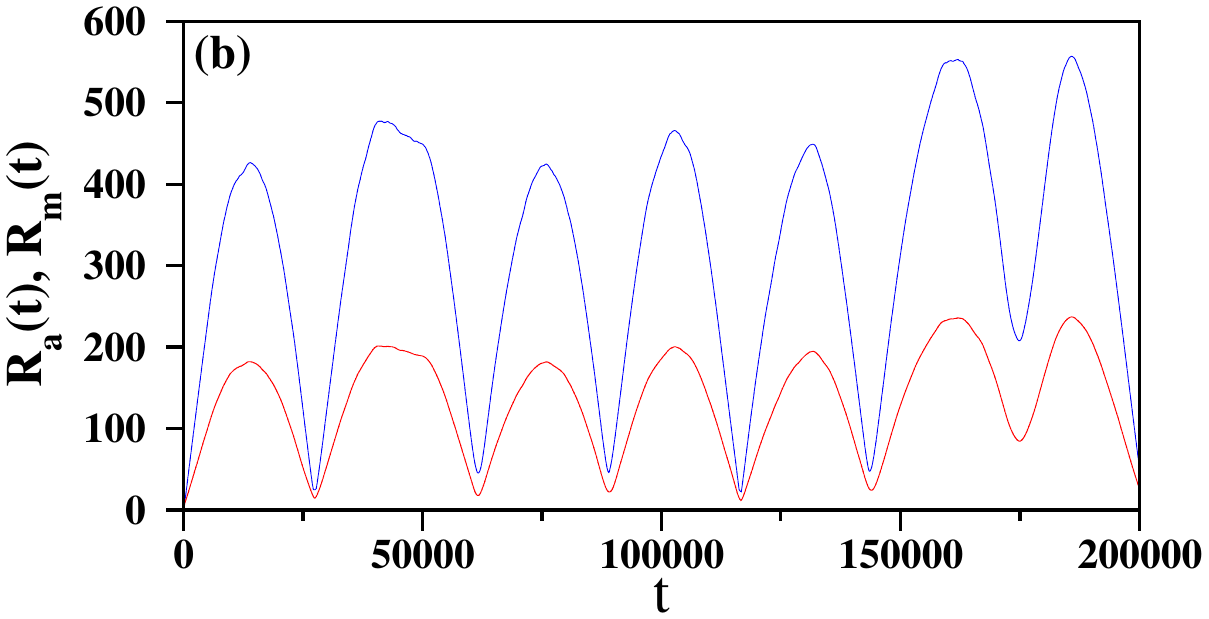} \\
\includegraphics[width=7.5cm]{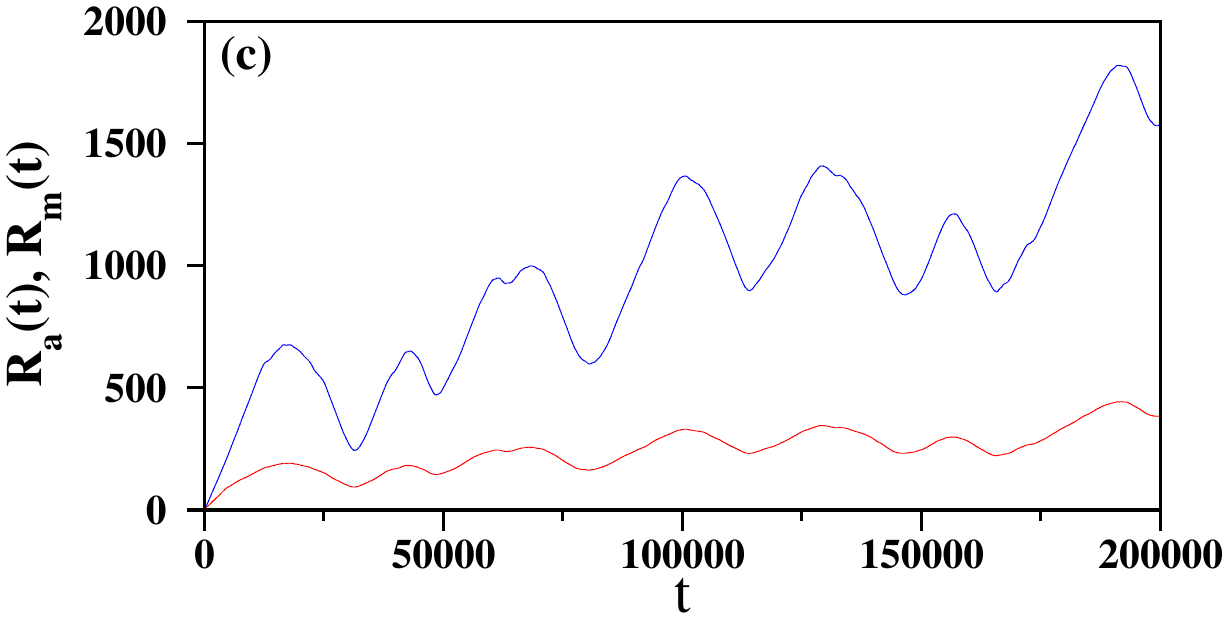} \hspace*{0.4 cm} & \includegraphics[width=7.5cm]{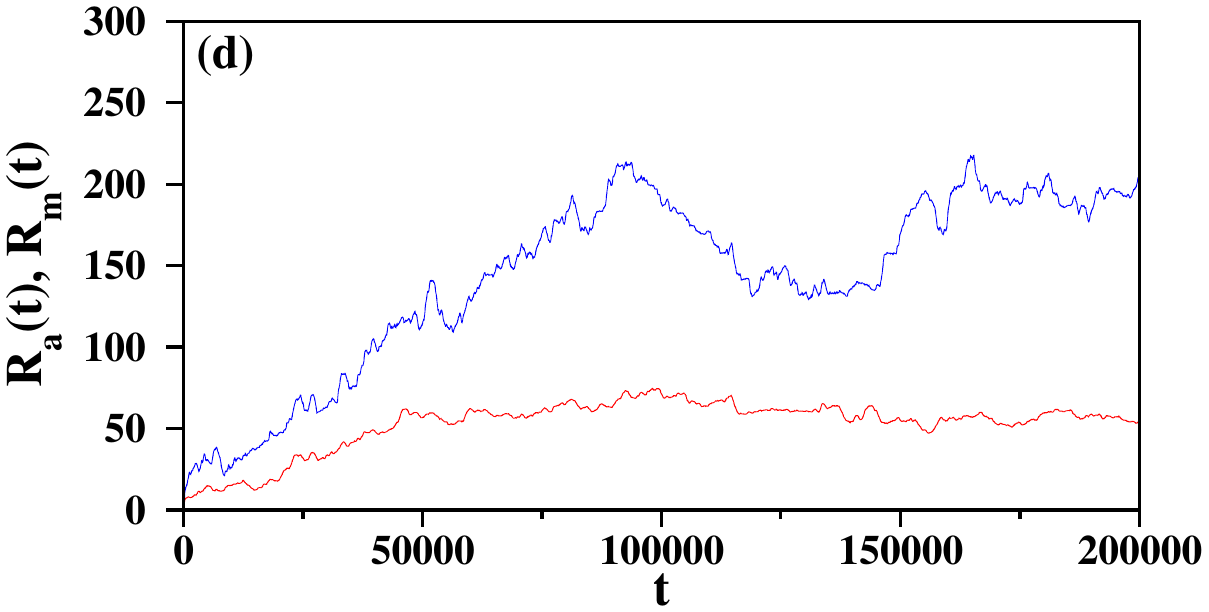} \\
\end {tabular}
\end{center}
\caption{(Color online)
Effect of noise is exhibited on a single flock ($N$ = 512 agents, each having $n$ = 10 neighbors) which 
goes to a cyclic state without any noise. Variation of the maximal radius $R_{m}$ (blue) and the average 
radius $R_{a}$ (red) have been shown with different strengths of the noise parameter: (a) $\eta = 0$, 
(b) 0.2, (c) 0.5 and (d) 1. The initial positions and velocities are same in all four cases.
}
\end{figure*}

\begin{figure}
\begin{center}
\includegraphics[width=7.0cm]{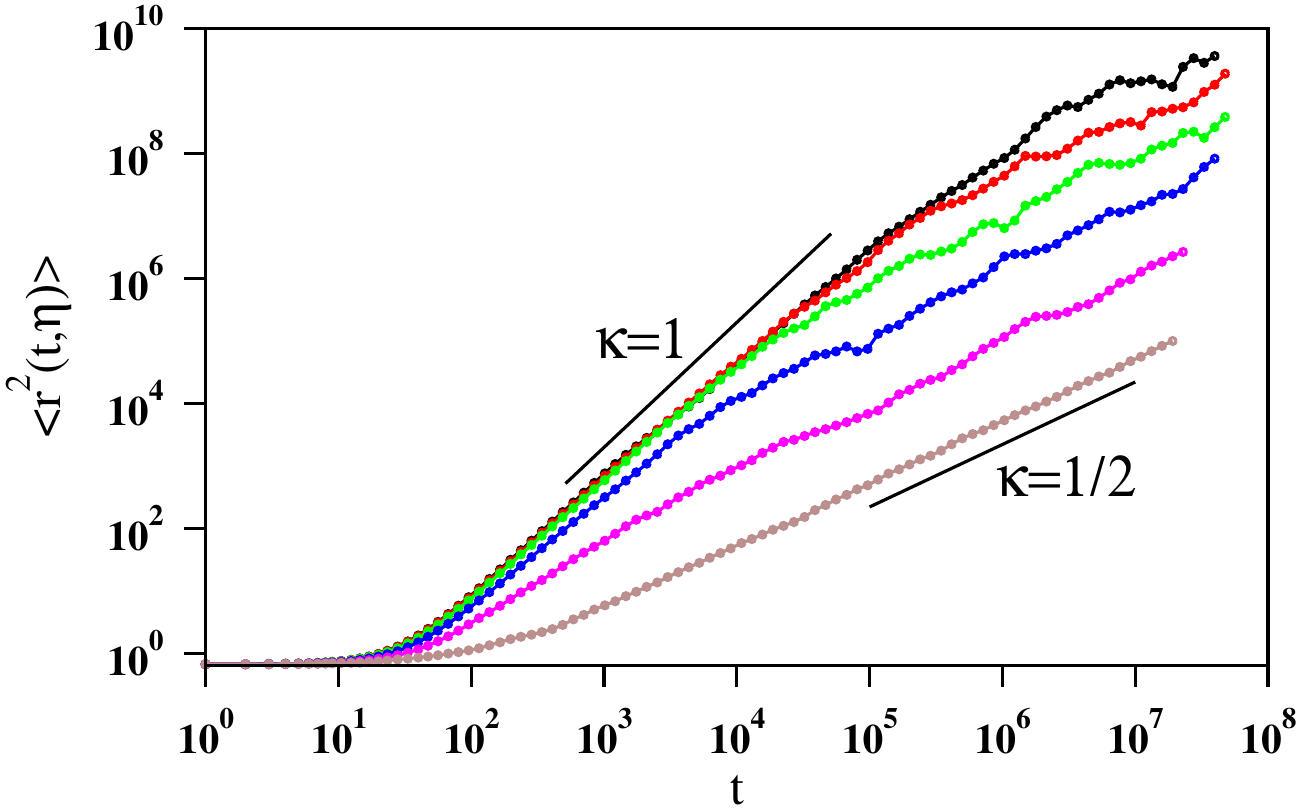}
\end{center}
\caption{(Color online)
The mean square displacement $\langle r^2(t,\eta) \rangle$ of an agent from the origin
has been plotted against time for different values of the noise parameter $\eta$ = 0.2 (black), 0.5 (red), 1.0 (green), 
2.0 (blue), 3.0 (magenta) and 4.0 (brown). The flock size $N=512$ and the neighbor number $n=10$. Two short straight lines
are the guides to the eye whose slopes are $\kappa = 1/2$ and 1.
}
\end{figure}

\begin{figure*}
\begin{center}
\begin {tabular}{ccc}
\includegraphics[width=5.5cm]{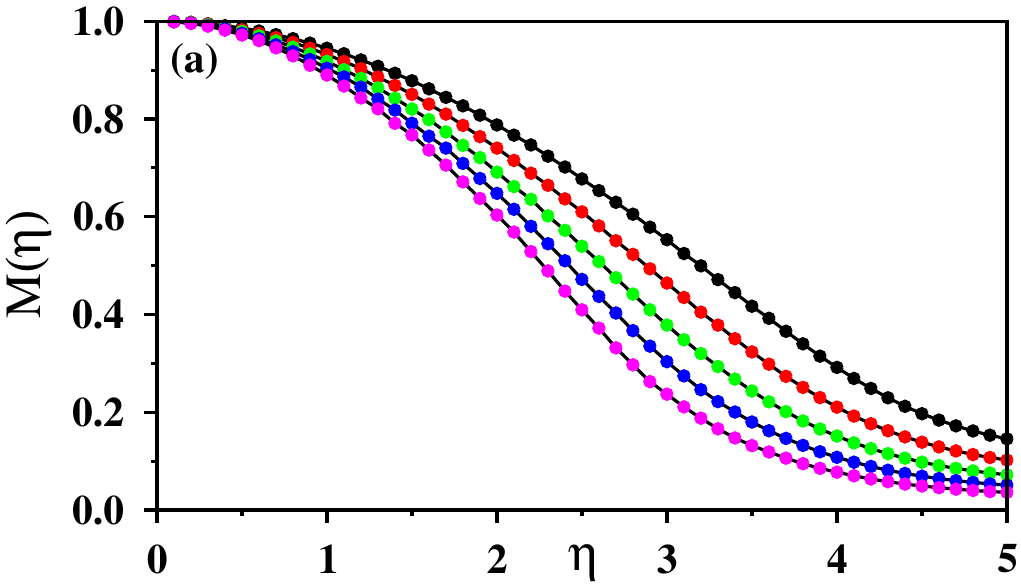} \hspace*{0.4cm}&
\includegraphics[width=5.5cm]{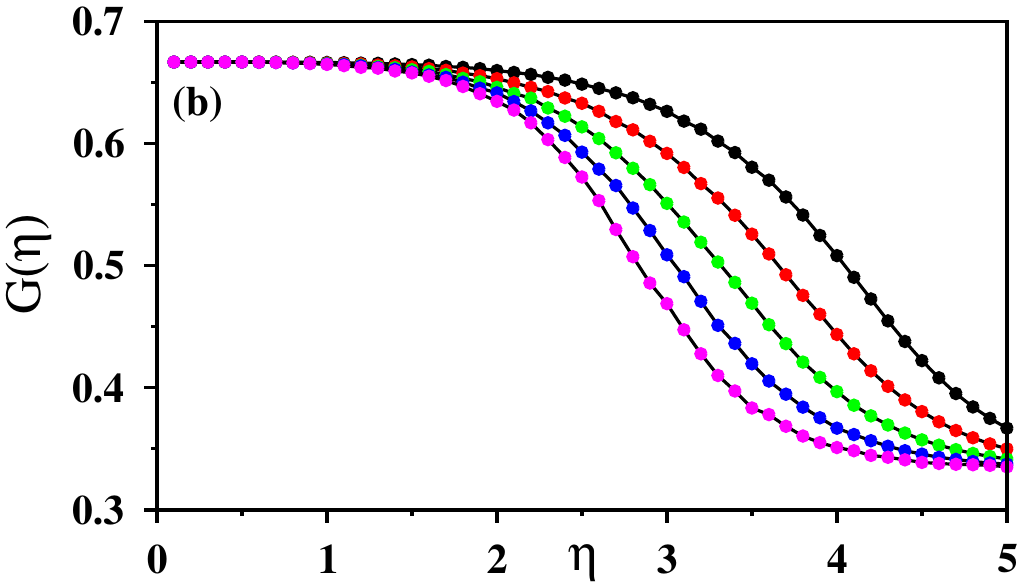} \hspace*{0.4cm}&
\includegraphics[width=5.5cm]{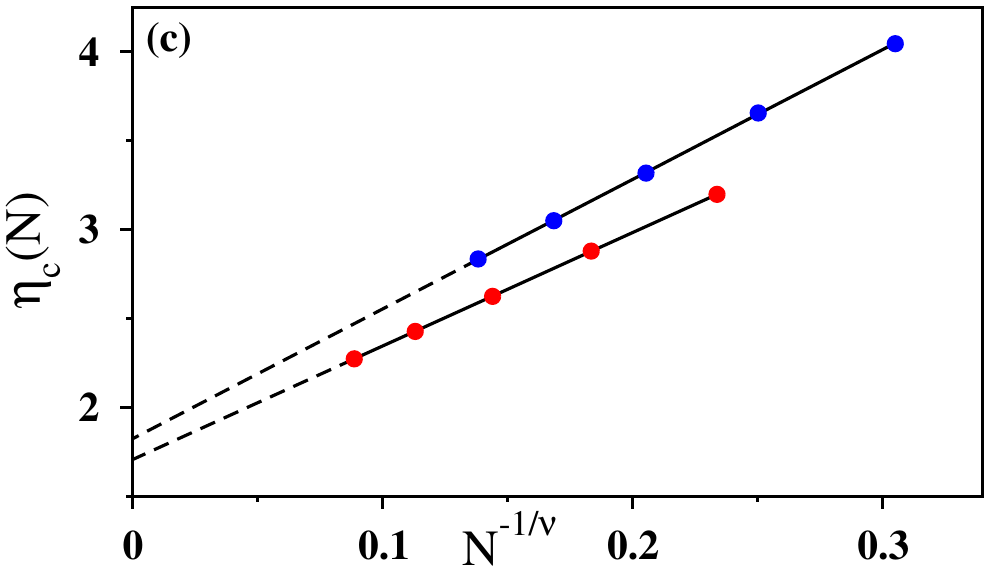} 
\end {tabular}
\end{center}
\caption{(Color online)
   (a) The stationary state order parameter $M(\eta)$ and 
   (b) the Binder cumulant $G(\eta)$ have been plotted for the system sizes $N$ = 64 (black), 128 (red), 256 (green), 
   512 (blue) and 1024 (magenta) where all agents start with their initial velocities in the same direction. System size 
   increases from right to left.
   (c) Extrapolation of $\eta_c(N)$ values determined from (a) and (b). We obtained $\eta_c(N) = 1.70 + N^{-1/2.86}$ and
   $\eta_c(N) = 1.82 + N^{-1/3.50}$ respectively.
}
\end{figure*}

\begin{figure}
\begin{center}
\includegraphics[width=7.0cm]{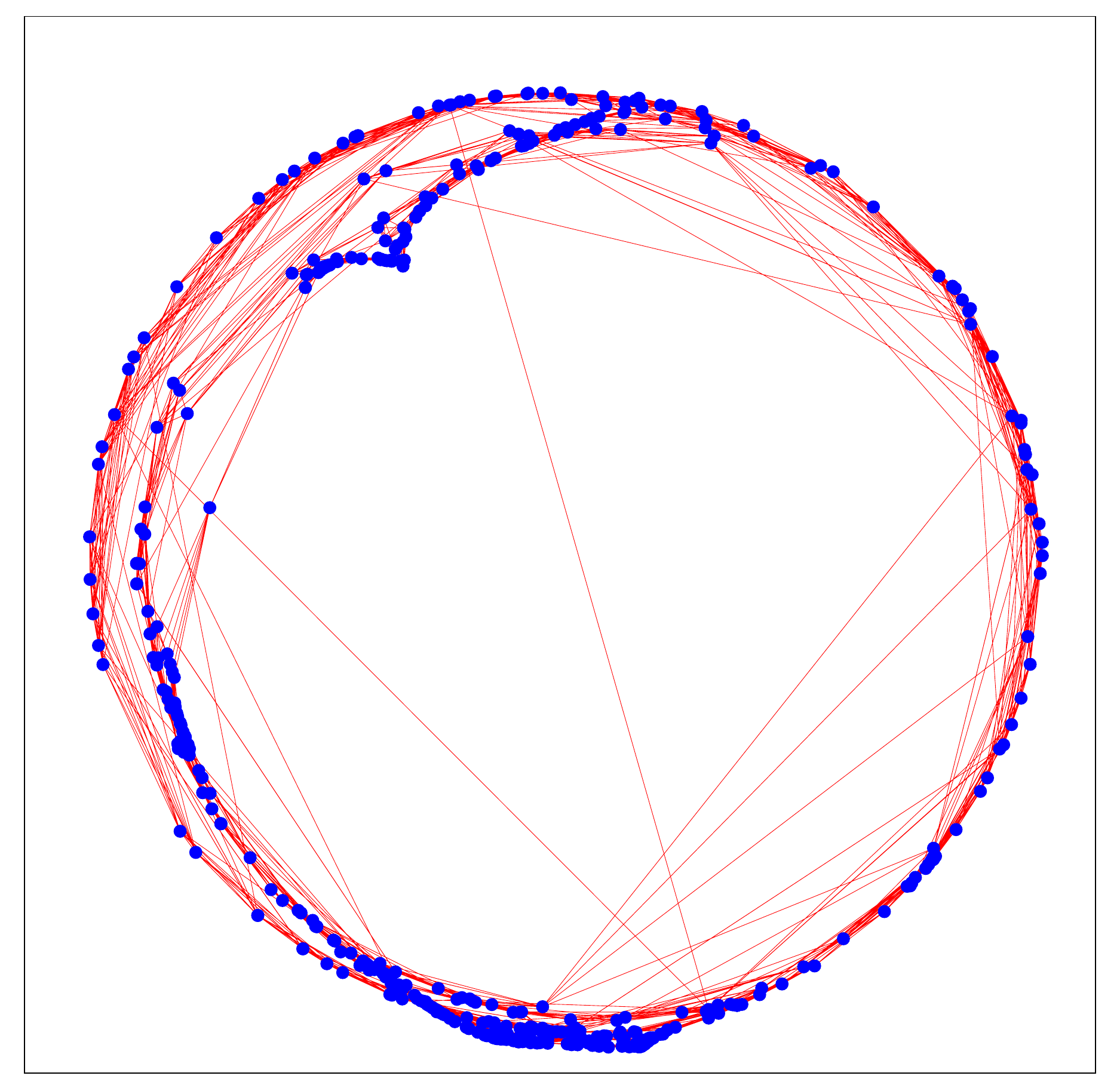}
\end{center}
\caption{(Color online)
The position of a flock ($N$ = 512, $n$ = 10) in CS by blue dots and links by red lines.
The frame size is 150 units.
}
\end{figure}

      Since the agents are uniformly distributed initially, the edges of RGG also are homogeneously distributed
   as shown in Fig. 2(d). However with time evolution these edges change their positions but their connectivity
   do not change, i.e., the end nodes of every edge are always fixed since the neighbor list does not change.
   How they look in the stationary state has been exhibited in Fig. 12. This is the picture of the circle shaped flock in the
   cyclic state. The blue dots represent the agents and the red lines represent the edges. What is interesting to note is
   that the system self-organizes itself so that not only agents but also the edges are constrained to be within a
   very limited region of the space. Very few edges criss-cross the flock from one side to the opposite side. Initially
   each agents had its $n$ neighbors at its closest distances. After passing through the relaxation stage and arriving at
   the stationary state, when the shape of the flock is completely different from its initial shape, most of the agents
   maintain their connections with other agents in their local neighborhood only.
   
\section{6. The fastest refreshing rate of the Interaction Zone}

      Here we consider the case corresponding to the fastest refreshing rate of the interaction zone, i.e., when 
   every agent updates its $n$ nearest neighbors at every time step. 
   Consequently, the RGG is no longer a constant of motion in this case and is updated at each time step. While performing 
   simulations of this version, we first notice that in the long time stationary state the entire flock becomes
   fragmented with probability one into different clusters. For a flock of $N$ agents with neighbor number $n$, the 
   minimum number of agents in a cluster is $n+1$. An agent of a particular cluster has all $n$ neighbors which are 
   members of that cluster only. In the stationary state all agents of a cluster has exactly the same direction of velocity 
   and the entire cluster moves along this direction with uniform speed $v$. Therefore, the velocity direction 
   $\theta$ of a particular cluster can be looked upon as the identification label of that cluster. The shape of the 
   flock is approximately circular since each cluster travels outward with same speed (Fig. 13(a)). Moreover, within a 
   cluster if the agent $i$ is a neighbor of the agent $j$ then $j$ is also a neighbor of $i$. Therefore the sub-graph 
   of the entire RGG specific to a cluster is completely undirected and the corresponding part of the adjacency matrix 
   is symmetric (Fig. 13(b)). This immediately implies that the $N \times N$ adjacency matrix of the entire flock can 
   be written in a block-diagonal form by assigning suitable identification labels of different agents.
   
\begin{figure*}
\begin{center}
\begin {tabular}{cc}
\includegraphics[width=7.0cm]{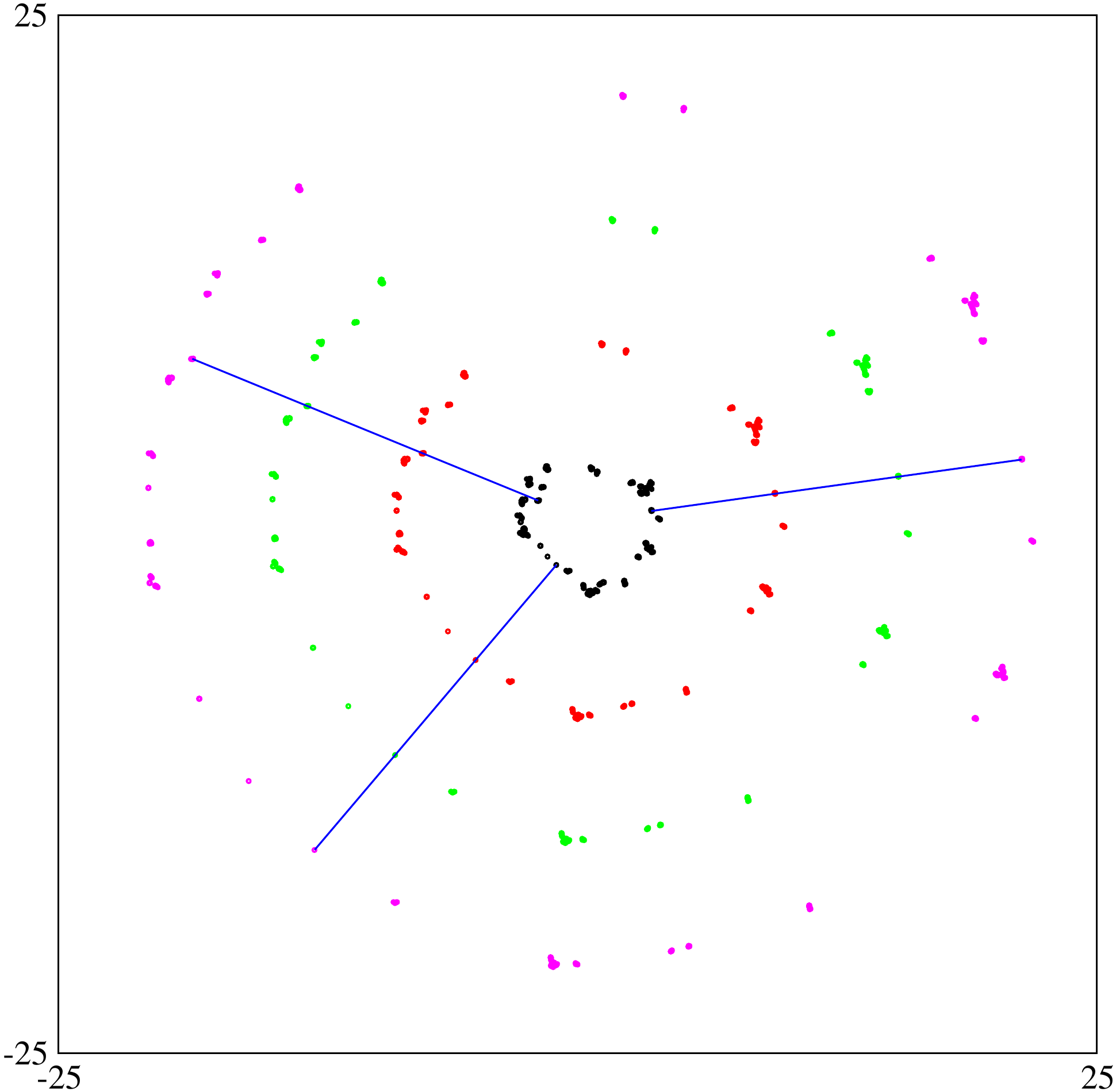} \hspace*{0.8cm}& \hspace*{0.8cm}
\includegraphics[width=7.0cm]{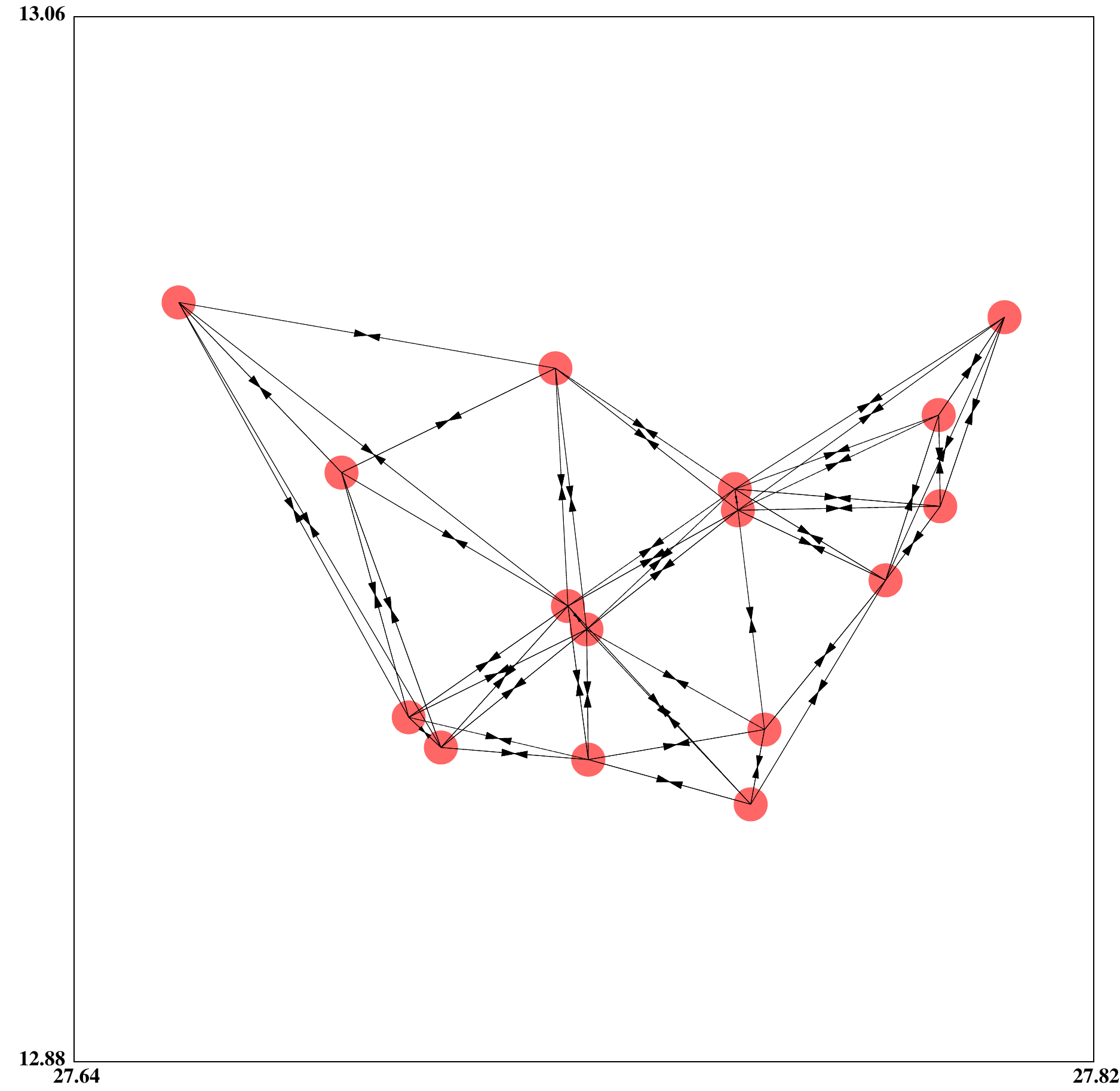} \\
{\bf (a)}  &  {\bf (b)}
\end {tabular}
\end{center}
\caption {(Color online) 
    (a) Positions of a flock of $N$ = 512 agents with $n=5$ neighbors in the interaction zone and at four different time 
    instants: 100 (black), 200 (red), 300 (green) and 400 (magenta). The RGG has been updated at every time step. Three 
    individual agents' straight line trajectories have also been shown. There are a total of 29 clusters.
    (b) The sub-graph of the RGG corresponding to a specific cluster of 16 agents have been shown. An arrow has been drawn
    from the agent $i$ to the agent $j$ if $j$ is one of the $n$ neighbors of $i$. There are a total of $16 \times 5$ = 80
    distinct links and each link is directed in both directions. It may also be noted that whole set of links are
    restricted to the nodes of the cluster only.
}
\end{figure*}

      A natural question would be how the probability distribution $D(s)$ of different cluster sizes depends on the cluster 
   size $s$. To answer this question a large number of independent flocks have been simulated and each of them was 
   evolved to its stationary state. In the stationary state the sizes of the individual clusters are measured using the 
   burning method. In Fig. 14(a) a plot of $D(s)$ vs. $s$ for $s > n$ has been shown on a semi-log scale for $N$ = 512 
   and $n$ = 5, the data being collected using a sample size of 20000 independent flocks. Apart from some noise at the 
   tail end and a maximum around the smallest value of $s$ the plot fits well to a straight line, implying an exponentially
   decaying form of the probability distribution. We conclude $D(s) \sim \exp(-s/s_c)$ where $s_c \approx 7.2(1)$.
     
      Next, the average number of clusters $\langle n_s(N,n) \rangle$ has been calculated and plotted in the inset of Fig. 14(b)
   for $N$ = 128, 256 and 512 and $n$ = 5 on a $\log - \log$ scale. Again, apart from the tail end, the plots fit very
   nicely to parallel straight lines, the slopes of which are estimated to be 1.168(5). In the main part of Fig. 14(b) a
   scaling has been shown which exhibits a nice data collapse, corresponding to the following form
\begin {equation}
\langle n_s(N,n) \rangle N^{-1.1} \sim n^{-1.168}.
\end {equation}
   This implies that as the neighbor number $n$ increases, there would be fewer clusters in the stationary state. 
   On the other hand, for a specific value of $n$, the average cluster number grows with the flock size as $N^{1.1}$. Assuming
   that the above scaling relation holds good for the entire range of $n$, for a given $N$ one can define a cut-off value 
   of $n = n_c$ such that $\langle n_s(N,n_c) \rangle$ = 1 which leads to $n_c(N) \sim N^{1.1/1.168} = N^{0.94}$. However our simulations
   suggest that due to the presence of an upward bending at the tail end, the above scaling relation does not work at this end
   and $n_c(N)$ is actually of the order of $N$.
   
\begin{figure*}
\begin{center}
\begin {tabular}{cc}
\includegraphics[width=7.0cm]{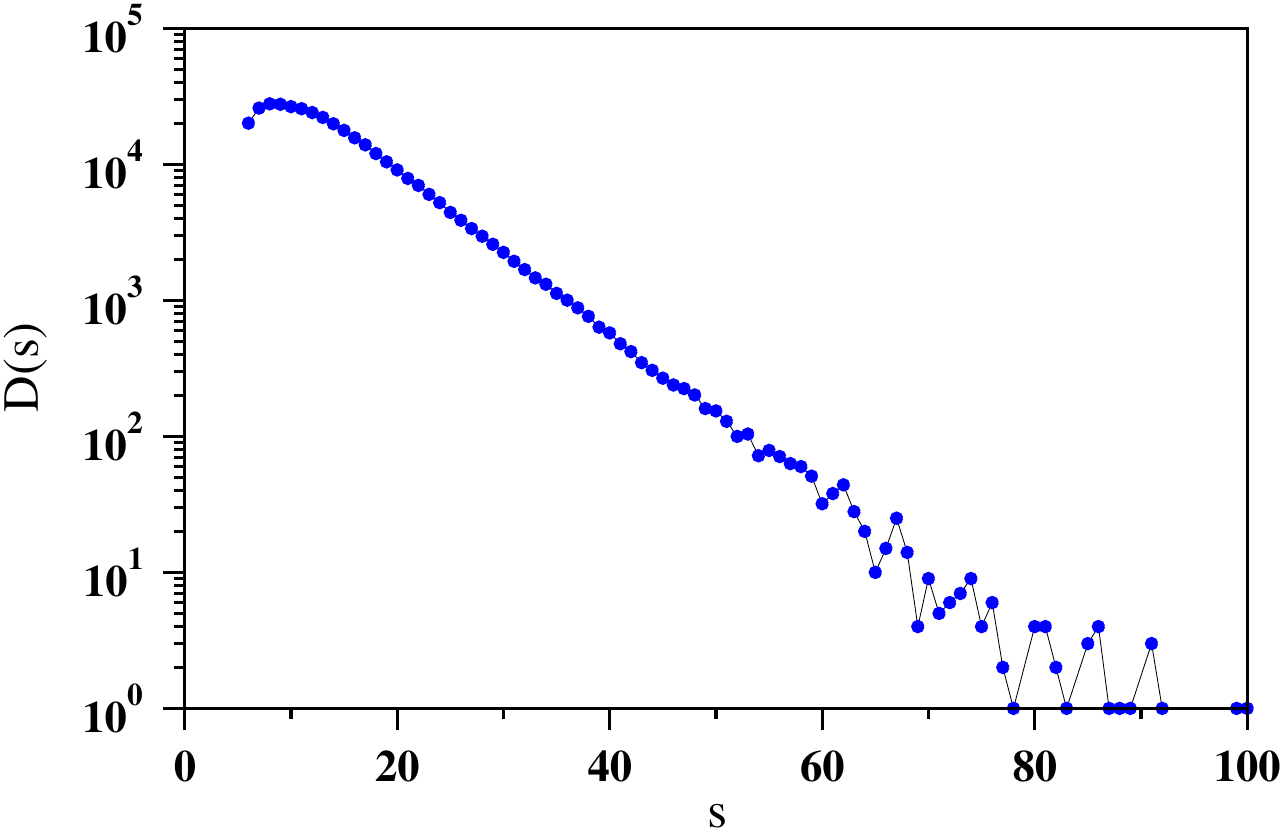} \hspace*{0.8cm}& \hspace*{0.8cm}
\includegraphics[width=7.0cm]{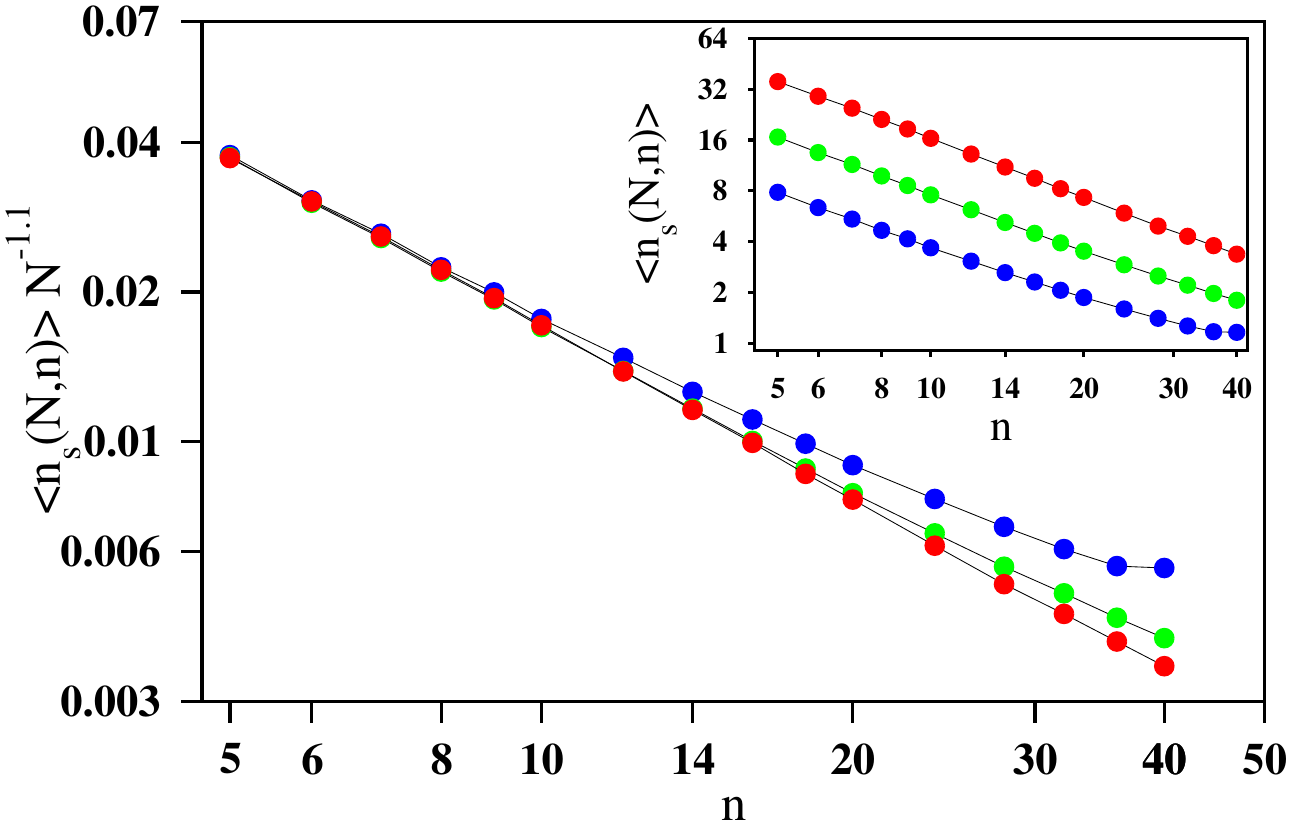} \\
{\bf (a)}  &  {\bf (b)}
\end {tabular}
\end{center}
\caption {(Color online) 
    (a) The probability distribution $D(s)$ of the cluster sizes $s$ seems likely to have an exponentially 
    decaying form $\exp(-s/s_c)$ where $s_c \approx 7.2$.
    (b) The inset shows the plot of the average cluster size $\langle s(N,n) \rangle$ for $N=512$ and $n=5$
    with neighbor number $n$ on a $\log - \log$ scale: $N$ = 128 (blue), 256 (green) and 512 (red). In the
    main plot the vertical axis has been scaled by $N^{1.1}$ which leads to a data collapse.
}
\end{figure*}
\vskip -0.5 cm
\section{7. Vortices on the square lattice}

      In this section we studied a simpler version of our model where every agent is a spin vector. They are no more
   mobile, their positions are completely quenched at the sites of a regular lattice, but the directions $\theta_i(t)$ 
   of the spins are the only dynamical variables that evolve with time following Eqn. (1). More specifically, spins 
   are placed on a square lattice with different choices for the first $n$ neighbors and we study the spatio-temporal 
   patterns that emerge during the time evolution of the angular variables $\{\theta_i(t)\}$. The arrangement of the 
   spins allows to draw a parallel with the dynamics of planar spins in the two dimensional $XY$ model.

      The connections between the Vicsek model \cite{Vicsek} and the $2d$ $XY$ model \cite{KT} have been explored since long 
   \cite{toner-tu,Chate-nematics,toner-tu-ram}. It is well known that in the limit of speed $v\to 0$ the dynamics of the Vicsek model would 
   exactly map on to the finite temperature Monte Carlo dynamics of the $2d$ $XY$ model. However, in the latter model 
   any long-range ordered phase is absent. Instead a quasi-long-range ordered phase appears at the low temperatures 
   and the transition to  the disordered phase is associated with the simultaneous unbinding and increase of 
   vortex-antivortex (VAV) pairs. In the low temperature phase VAV pairs are to be found in tightly bound states. 
   We find that our model defined on the square lattice also gives rise to VAV pairs and we determine the density 
   of such pairs, as a function of the noise amplitude $\eta$.

      We define the IZ of an agent with respect to its $n$ nearest neighbors on the square lattice of size $L \times L$ with
   the periodic boundary condition. For $n=2$, the IZ includes the top and the right nearest neighbors. For $n=3$, the left 
   nearest neighbor is also included and in the case of $n=4$, all the four nearest neighbors are included. We notice that
   for the $n=4$ case, the Vicsek model with spins similarly placed on the square lattice and interact with a range $R = 1$ 
   and our model are same. As before we study the dynamics of the spin system with and without noise.

\begin{figure*}
\begin{center}
\begin {tabular}{cc}
\includegraphics[width=7.0cm]{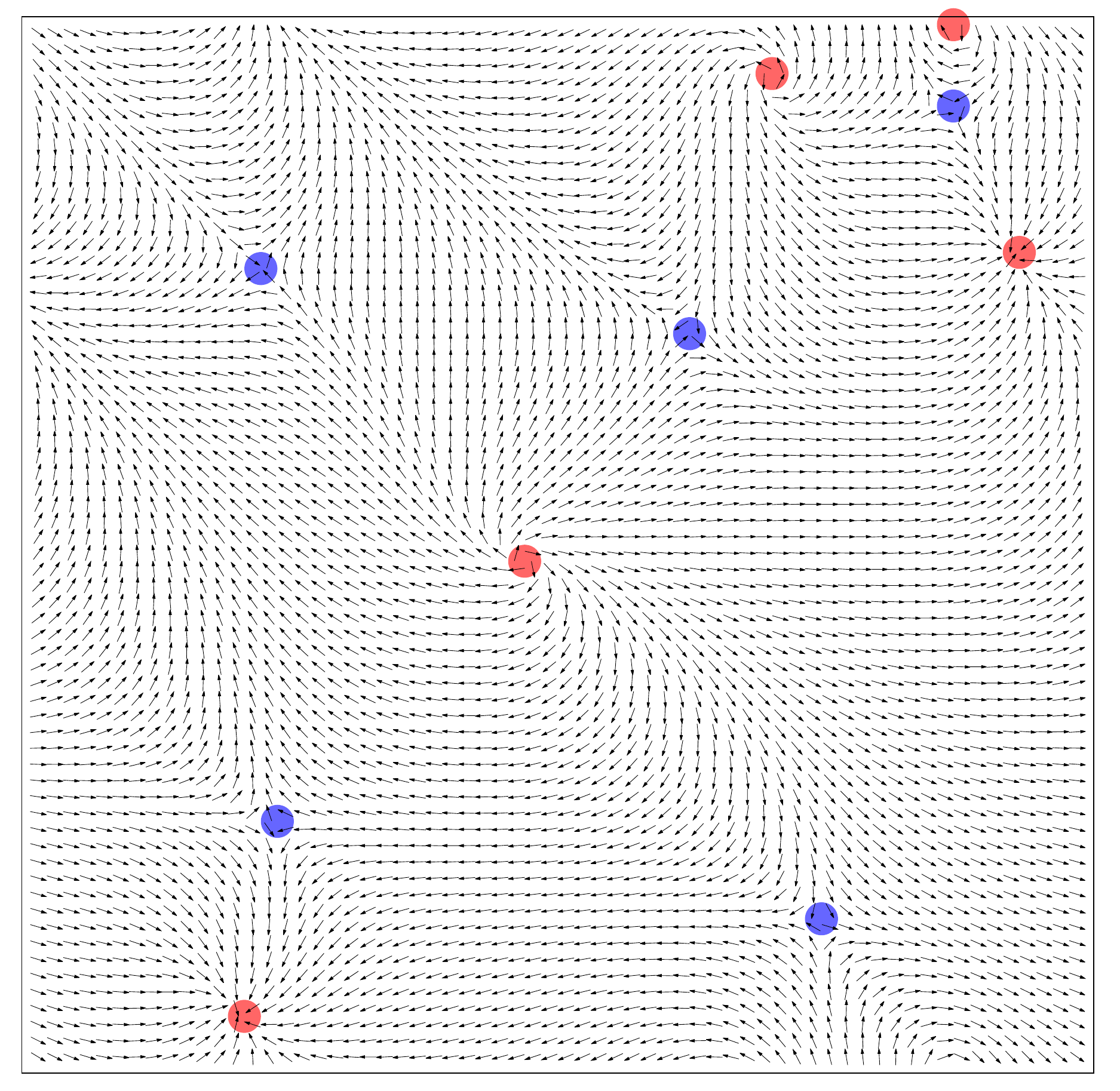} \hspace*{0.8cm}& \hspace*{0.8cm}
\includegraphics[width=7.0cm]{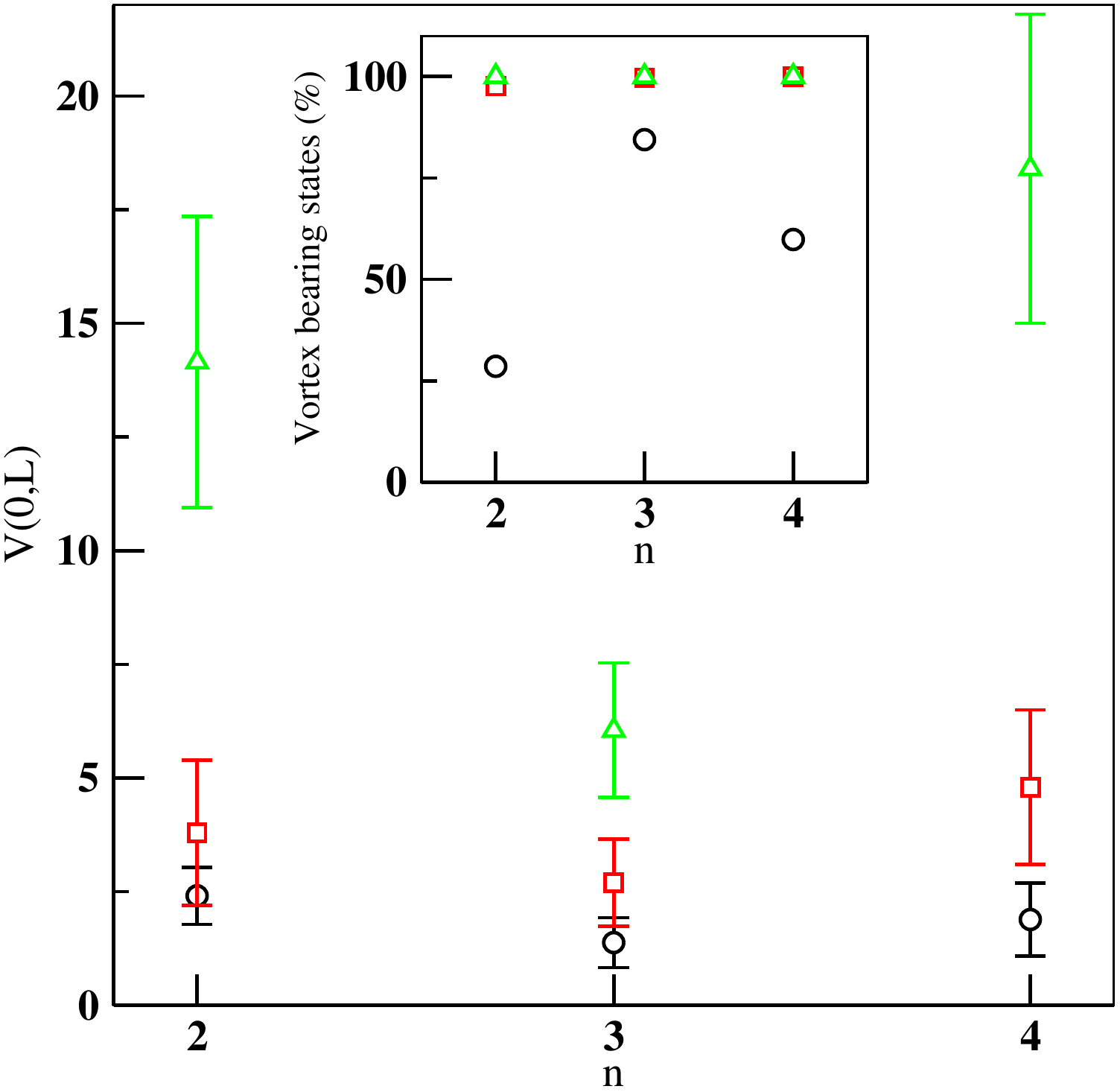} \\
{\bf (a)} \hspace*{0.8cm}& \hspace*{0.8cm} {\bf (b)} 
\end {tabular}
\end{center}
\caption {(Color online)
   (a) Vortex-antivortex pairs of the spin systems in the stationary state in a square lattice with size $L=64$ and 
   $n=2$ with zero noise. The orientations of the spins change with time in such a way so that the entire spin pattern 
   with vortices and the antivortices move with a uniform speed from the top-right corner to the 
   bottom-left corner. Vortices and antivortices are marked by filled red and blue circles. 
   (b) The number $V(0, L)$ of vortex-antivortex pairs at zero noise for different values of $n$ in lattices of three 
   different sizes: $L=32$ (circles), $64$ (squares) and $128$ (triangles). The bars indicate the standard deviation
   in the values obtained from around $100$ configurations in each case. In the inset the percentage of configurations 
   which lead to steady states with vortex pairs is plotted against $n$.}
\vskip 0.4 cm
   \end{figure*}

      In the absence of noise, beginning from arbitrary initial conditions for $\theta_i$'s, the dynamics results 
   in the formation of VAV pairs. For $n=2$ and 3, the interactions are anisotropic. Consequently the entire spin
   pattern in the stationary state as well as all VAV pairs are mobile and in general all the spin orientations 
   $\theta_i$'s change with time. In comparison, for $n=4$, all the $\theta_i$'s remain frozen in time which also
   implies that all the VAV pairs are anchored.  We find that the choice of the IZ, in addition to the periodic
   boundary condition, fixes the direction of motion of the VAV pairs. In the Fig. 15(a) an instantaneous configuration 
   of the spins is plotted for a lattice with $L=64$ and $n=2$.  In general for the $n=2$ case the entire spin 
   pattern moves on the average along the diagonal direction from top-right to bottom-left. For the case, $n=3$, 
   the vortices travel from the bottom to the top. 

      Let the $V(\eta,L)$ be the number of VAV pairs observed at noise $\eta$ in lattice of size $L$. The number 
   of VAV pairs observed at zero noise, $V(0,L)$ is plotted against the value of $n$ in the Fig. 15(b) for different 
   lattice sizes. We observe during the time evolution for a given initial condition that the number 
   of VAV pairs initially decays and then becomes stationary. However, different initial conditions leads to different 
   values at the stationary states. The bars indicate the dispersion that is observed for different initial conditions 
   which lead to non-zero number of VAV pairs. For the lattice size $L=128$ we  wait for 
   $10^5$ steps before calculating number of VAV pairs. The inset to the Fig. 15(b) shows the percentage
   of initial conditions that lead to non-zero number of VAV pairs. We find as lattice size increases, arbitrary initial 
   conditions, almost always, lead to states with VAV pairs.   

\begin{figure*}
\begin{center}
\begin {tabular}{ccc}
\includegraphics[width=7.5cm]{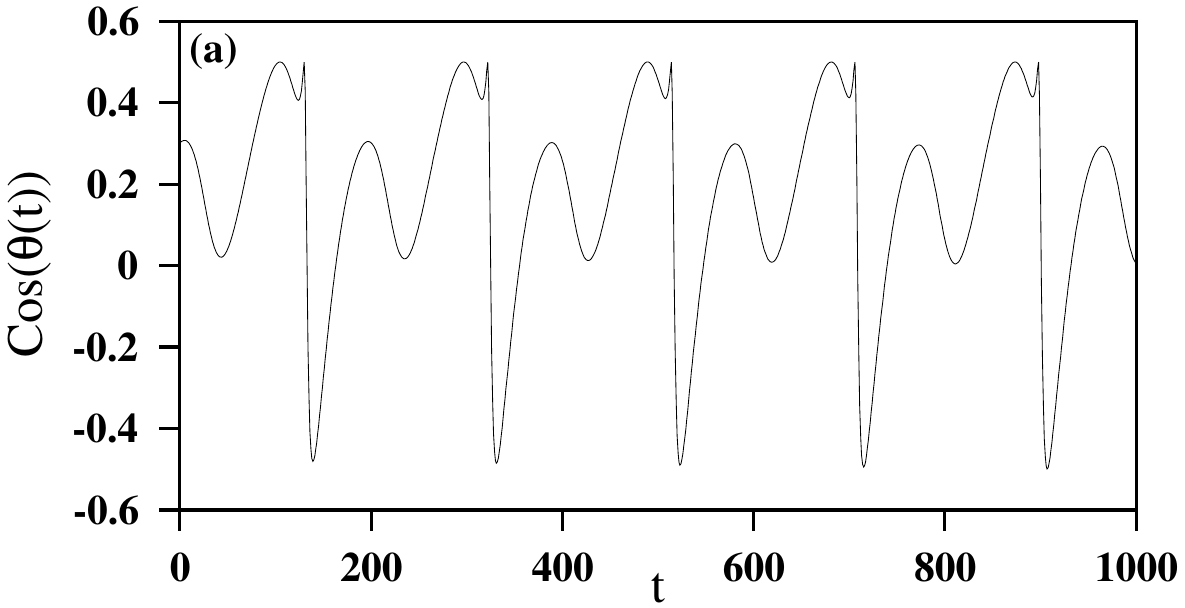} \hspace*{0.2cm}& \hspace*{0.2cm}
\includegraphics[width=7.5cm]{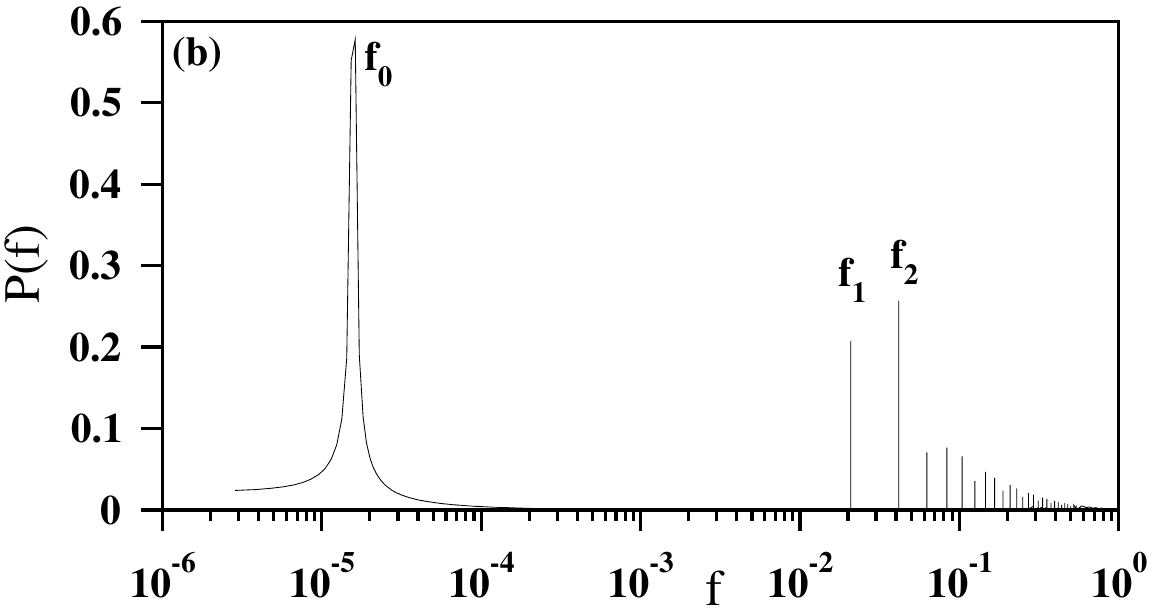}
\end {tabular}
\end{center}
\caption{(Color online)
(a) The $x$-component of a typical spin vector is plotted against time for lattice with $L=64$ and $n=2$
at zero noise. (b) The corresponding power spectrum indicates the 
presence of three basic frequencies $f_0=1/(4096L)$, $f_1=1/(3L)$ and $f_2=2/(3L)$. The higher frequencies can be expressed
as combinations of the $f_1$ and $f_2$ and essentially become harmonics of $f_1$ since $f_2=2f_1$.}
\end{figure*}

      The nature of the time variation of the spin angle $\theta_i$'s for the cases $n=2$ and $n=3$ are found 
   to be quite complex. In the Fig. 16(a) we plot the time series corresponding to the oscillation of the 
   $x$-component of a typical spin vector for $L=64$ and $n=2$. The corresponding power spectrum apparently 
   reveals the presence of three basic frequencies $f_0$, $f_1$ and $f_2$ all of which are rational multiples 
   of $1/L$ (Fig. 16(b)). The time evolution can be explained as a periodic oscillation with  $f_1$ and $f_2$ 
   (since $f_2=2f_1$) riding on a very slow mode.  These features carry over to $L=128$ as well. We find that 
   in the case of $n=3$, the spectrum is similar but the frequencies are not simple multiples of $1/L$. 

      It is known that metastable vortices are produced at low temperatures in the $2d$ $XY$ model when equilibrium is achieved 
   beginning from arbitrary initial conditions. However, these vortices are not responsible for the VAV unbinding transition
   \cite{miyashita}. Therefore, we study the effect of noise by ``cooling down'' \cite{tobochnik} the system to zero noise level 
   starting from a high value of noise. At each noise level the system initially passes through $10^5$ time steps; after 
   this relaxation it passes through an additional $10^4$ time steps and then moves to the next lower level of noise.
   We begin around the value of noise given by $\eta=3.7$ and decrease $\eta$ by an amount $0.07$ in each step. This method 
   suppresses the generation of the VAV pairs at low noise. At zero noise the vortices are absent in contrast to the statistics 
   discussed in the previous paragraphs where the cooling down method was not employed. 

\begin{figure*}
\begin{center}
\begin {tabular}{ccc}
\includegraphics[width=5.5cm]{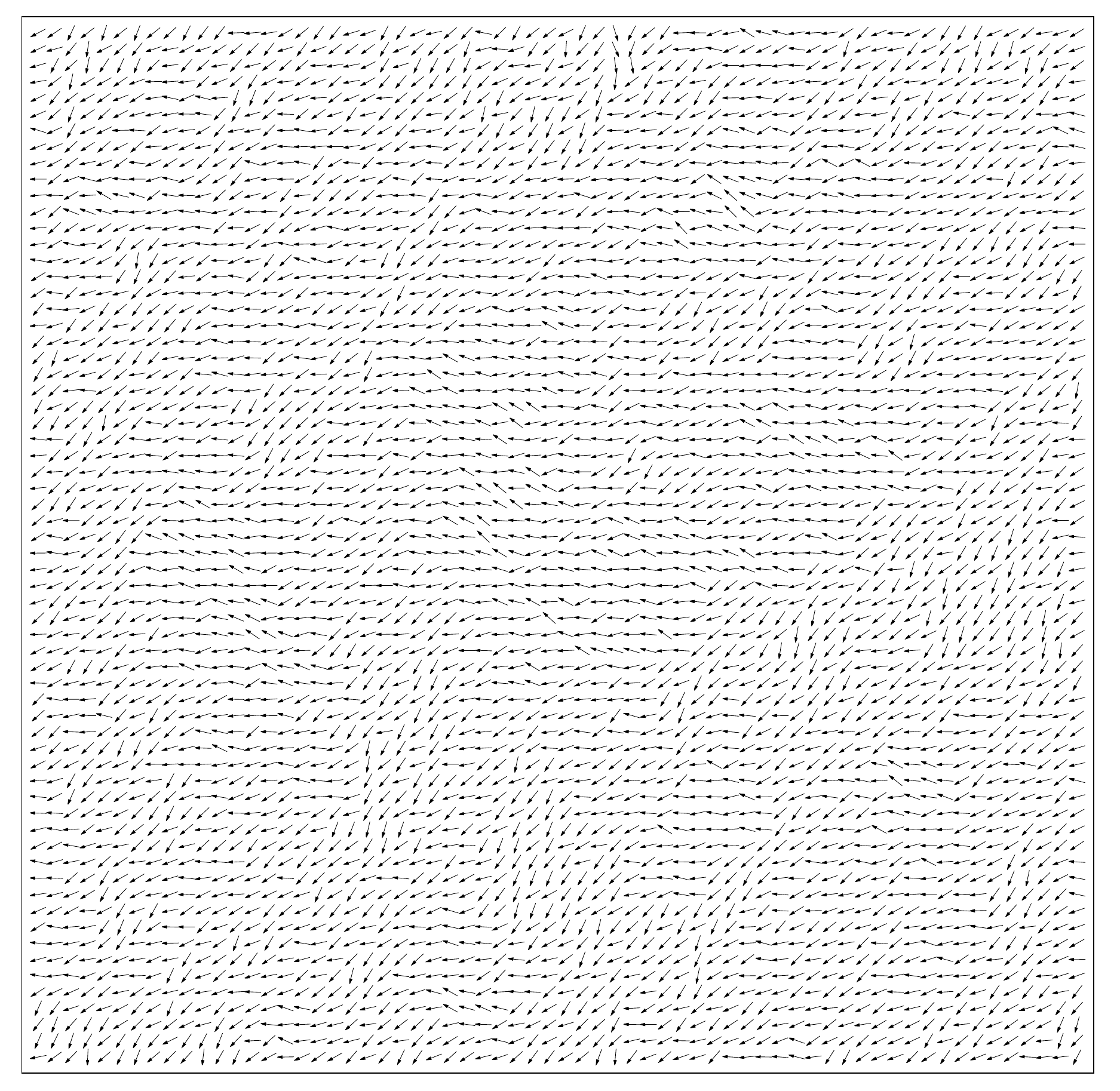} \hspace*{0.2cm}& \hspace*{0.2cm}
\includegraphics[width=5.5cm]{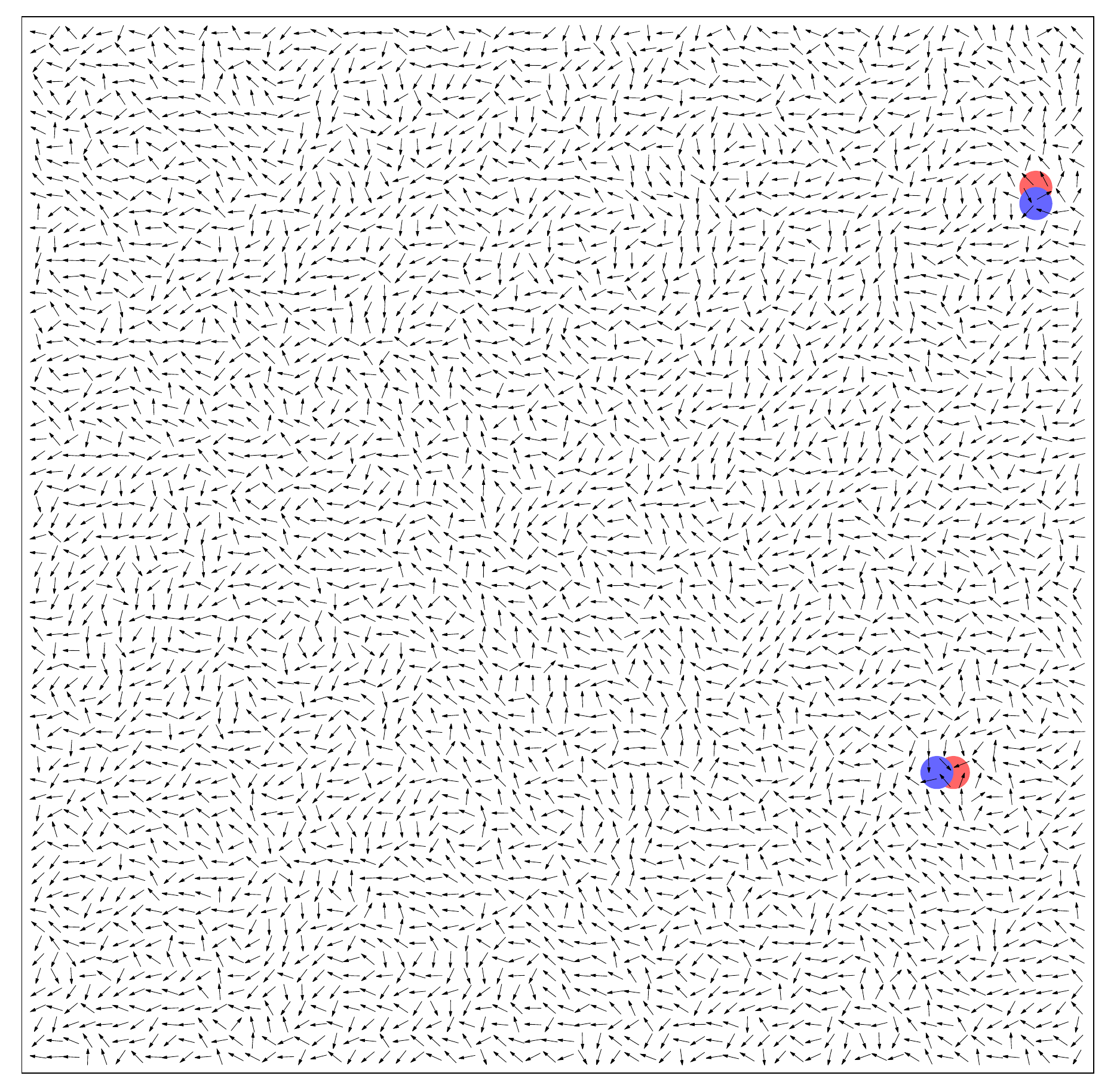} \hspace*{0.2cm}& \hspace*{0.2cm}
\includegraphics[width=5.5cm]{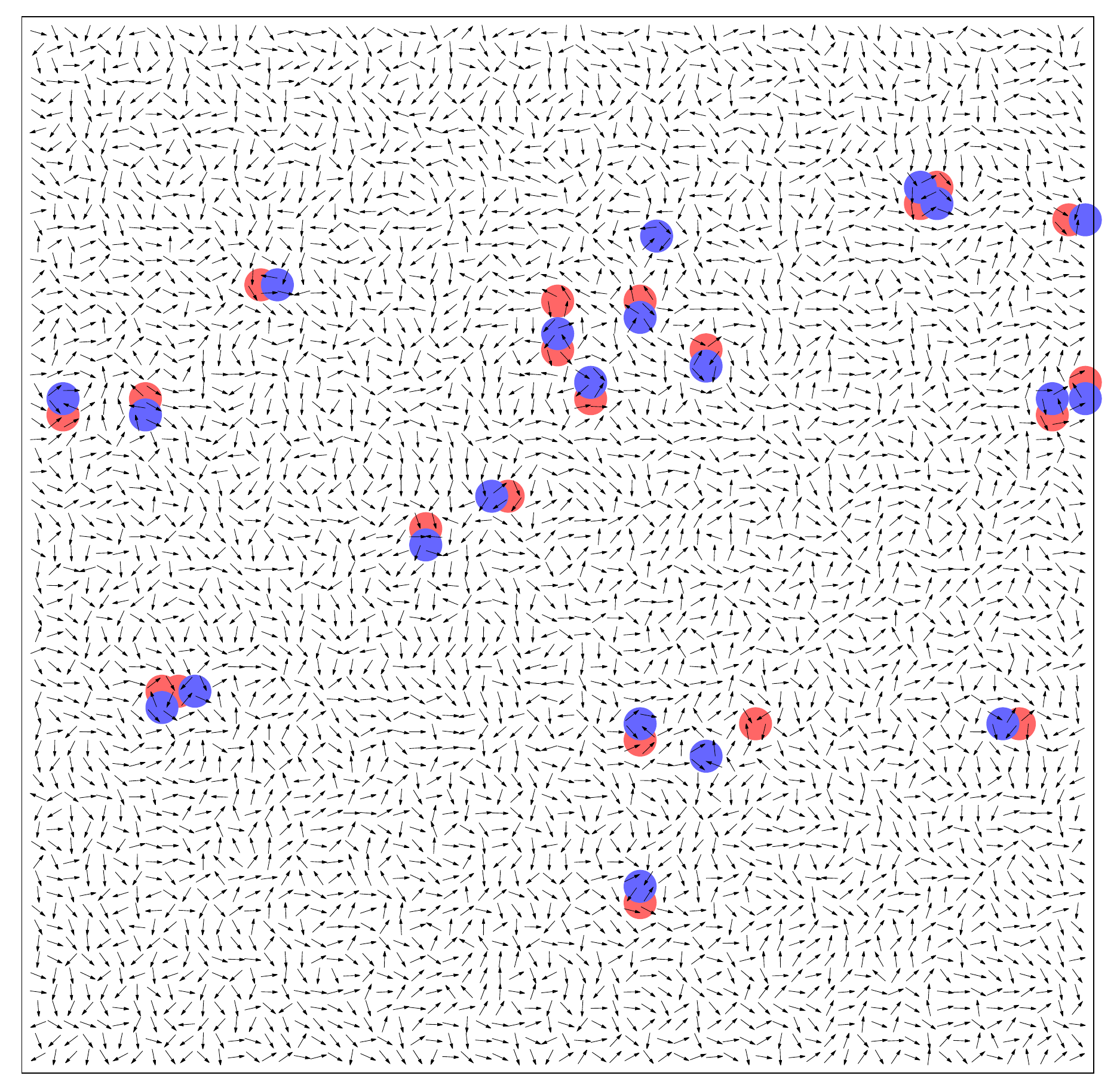} \\
{\bf (a)} \hspace*{0.2cm}& \hspace*{0.2cm} {\bf (b)} \hspace*{0.2cm}& \hspace*{0.2cm} {\bf (c)}
\end {tabular}
\end{center}
\caption {(Color online) 
The figure shows proliferation of vortices as noise is increased in a square lattice of length $L=64$ with $n=4$.
(a) For $\eta=0.15$ there are no vortices, (b) for $\eta = 0.35$ vortices begin to appear, there are only 
two vortex-antivortex pairs and (c) for $\eta = 0.40$ twenty one vortex-antivortex pairs can be seen. Filled
circles of two different colors, red for vortices and blue for antivortices, have been drawn around the vortex centers.} 
\end{figure*}

      At zero noise we find the spin system reaches the globally ordered state where all the spin vectors are oriented 
   in the same direction. We believe 
   that this is due to the finite size effect of the lattice. The spin configuration at low noise of $\eta = 0.15$ has been 
   shown in the Fig. 17(a) for $L = 64$ and $n = 4$ where not only vortices are absent but the long-range order is also not 
   present. At the higher noise levels VAV pairs start appearing and there is a rapid increase in the number of pairs with 
   further increase in noise. All the VAV pairs appearing initially are tightly bound i.e., lattice spacing is small between 
   the members in a pair as in Fig. 17(b)) but at higher noise members in a pair are seen to unbound (Fig. 17(c)). 
   
      The order parameter $M(\eta)$ tends to unity as $\eta \to 0$ as shown in the Fig. 18(a). In Fig.18(b) 
   we plot the vortex-pair density, defined as $\rho=V/L^2$. The natural collapse of the plots belonging to different 
   system sizes indicates a functional dependence of $\rho$ on $n$ and $\eta$ independent of $L$. To understand this 
   behavior we obtain the collapse of the logarithm of $1/\rho$ in Fig.18(c) for different values of $L$ and $n$. The 
   plot reveals that in the region where VAV pairs start proliferating $\rho \sim \exp(-\frac{An^{\beta}}{\eta^\alpha})$, 
   where $A$, $\beta$ and $\alpha$ are constants. We estimate $\alpha$ by averaging slope of individual curves which yields 
   $\alpha=2.68\pm 0.13$. This gives $\beta=\gamma \alpha=0.75$. This result is in contrast to the relation 
   $\rho \sim \exp(-\frac{a}{T})$, where $a$ is VAV pair energy and $T$ is the temperature for the $2d$ $XY$ model 
   \cite{tobochnik}.

\begin{figure*}
\begin{center}
\begin {tabular}{ccc}
\includegraphics[width=5.5cm]{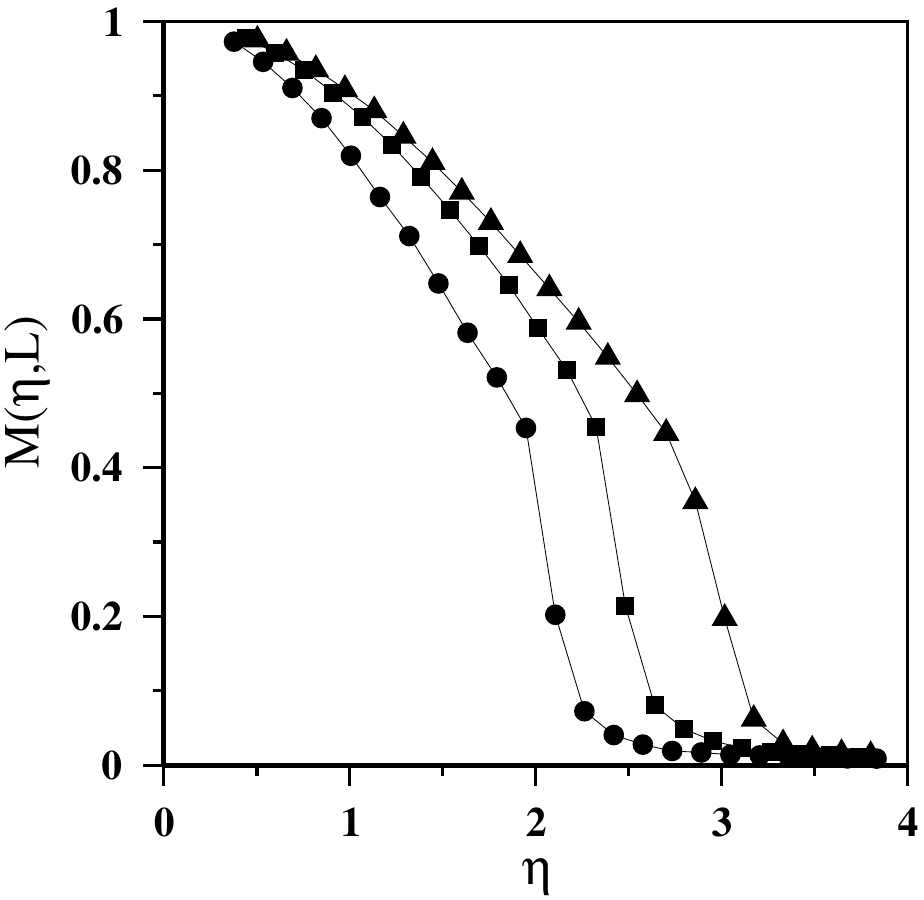} \hspace*{0.4cm}&
\includegraphics[width=5.5cm]{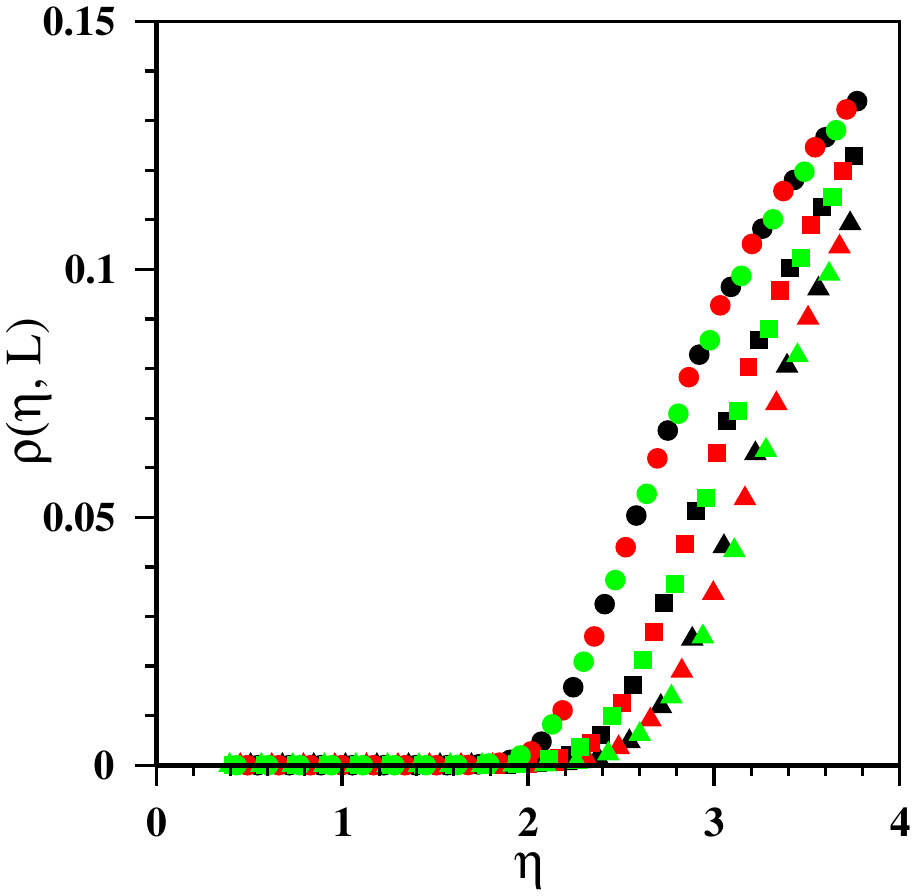} \hspace*{0.4cm}&
\includegraphics[width=5.5cm]{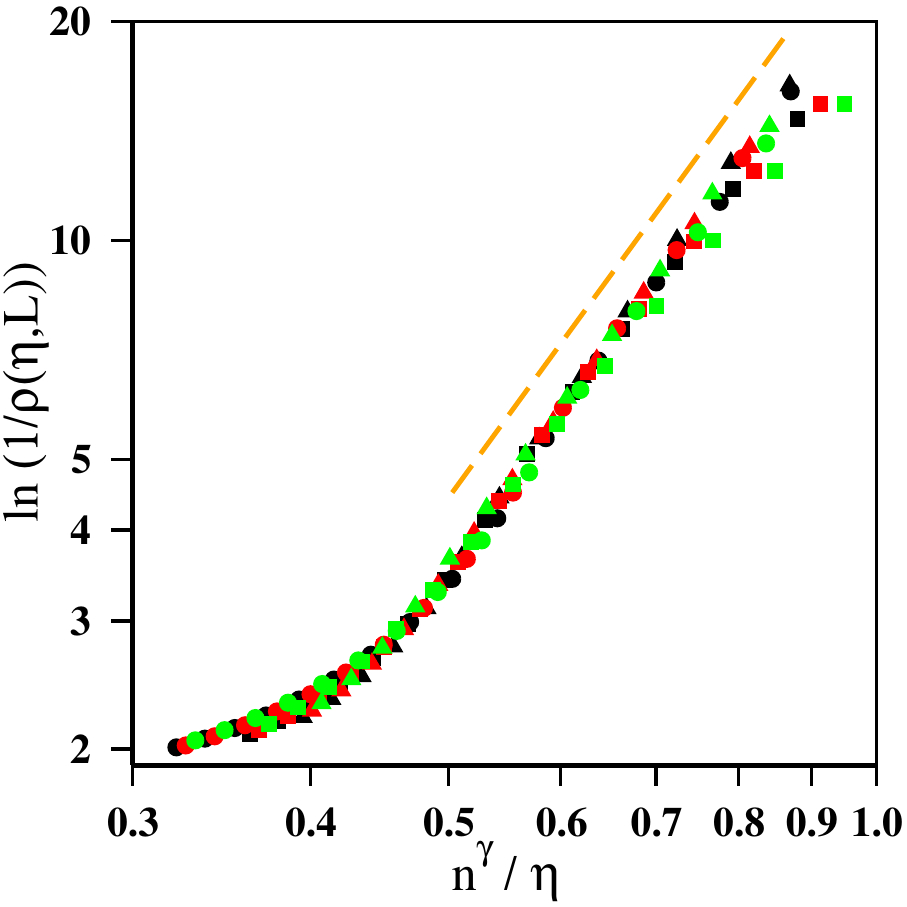} \\
{\bf (a)} & {\bf (b)} & {\bf (c)}
\end {tabular}
\end{center}
\caption{(Color online) The behavior of different quantities against noise for lattice sizes L = 32 (black), 64 (red) 
and 128 (green) with values of $n$ = 2 (circle), 3 (square) and 4 (triangle). 
(a) The variation of the order parameter $M(\eta, L)$ against noise $\eta$ are shown for $L = 128$ and
for $n = 2$, 3 and 4. 
(b) The vortex-pair density $\rho(\eta, L)$ against $\eta$. 
(c) Scaling collapse of $\ln (1/\rho(\eta, L))$ against $n^{\gamma} / \eta$. For the collapse 
we use $\gamma =0.28$. The dashed straight line, having slope $\alpha$ = 2.68, is a
guide for the eye and indicates the power-law nature in the low noise regime.} 
\end{figure*}
\vskip -1.5 cm
\section{8. Summary and Discussion}

      To summarize, we have studied a simple model of the collective behavior of $N$ interacting mobile agents which travel 
   in the open free space. In this model we have incorporated the observation of the StarFlag experiment which advocates
   the necessity of using the topological distance instead of the metric distance. Each agent interacts with a group of $n$ 
   selected agents around it who are positioned 
   within an interaction zone. Every agent freshly selects at a certain rate the group of agents within the IZ. 
   The selection criterion is to choose the first $n$ neighbors. In this paper we have studied two limiting 
   situations, i.e., when the refreshing rates are fastest and the slowest. We first studied when the refreshing rate
   is slowest i.e., when the IZ for each agent is determined at the beginning and is never updated.
   All the agents follow the interaction rule in Vicsek model. It has been observed that in absence of noise, starting 
   from a small localized region of space the agents gradually spread as time passes, so that after some relaxation 
   time the flock arrives at a stationary state. The most prominent stationary states are the single sink state and 
   the cyclic state. Using numerical methods we claim that the frequencies of occurrence of other stationary states 
   like the distributed sink states, cycloid states and the space-filling states goes to zero as the neighbor number 
   increases to about 8. Beyond $n=8$ only the SSS and CS states dominate. Finally as $n$ approaches the flock size 
   $N$, only the SSS states dominate. 
   
      Interestingly, it has also been observed that the actual metric distances of the $n$ topological neighbors do
   not become arbitrarily large in all these stationary states in absence of noise. For the SSS states it is true 
   exactly. For the Cyclic states also the topological neighbors remain within a finite distance from an agent (see
   Fig. 12). Moreover, when the refreshing rate is the fastest, every fragmented cluster travels in a single
   sink state. Therefore, it is this close proximity of the neighbors that gives rise to the cohesiveness present
   in our model.
   
      Further, on the application of noise a crossover takes place from the ballistic 
   motion to the diffusive motion and the crossover time depends on the strength of the noise $\eta$, which diverges 
   as $\eta \to 0$. Further the calculation of the Order Parameter $M(\eta)$ and the Binder cumulant $G(\eta)$ lead 
   us to estimate the critical noise $\eta_c$ required for the continuous transition from the ordered to the 
   disordered phase. Secondly, when the refreshing rate is fastest, each agent freshly determines its neighbors
   in the interaction zone at every time step. In the stationary state the flock gets fragmented into a number of
   smaller clusters of different sizes. The agents in a cluster move completely coherently, different cluster has 
   different direction of motion.
   
      A simpler version of the model has also been studied in the limit of the speed $v \to 0$ when the positions of 
   spins are completely frozen at the sites of a square lattice,
   but their orientational angles $\theta_i(t)$ evolve with time again by the Vicsek interaction. Here for $n$ = 4 the
   spin configuration is completely static. On the other hand for $n$ = 2 and 3, the entire spin configuration moves
   along the diagonal and parallel to the asymmetry axis respectively. Further we have observed that the density of 
   vortex-antivortex pairs increases with the strength of the noise and fits to a nice finite-size scaling behavior. 
   
      Overall, our findings suggest that complex spatio-temporal patterns may emerge in the interplay between an 
   underlying network structure and collective motion. We believe that our study would also be relevant in the 
   general problem of consensus development in networked agents \cite{Olfati-Saber} and as such issues like undesired 
   synchronization observed in real-world networks \cite{sally-floyd}. We observed that multiple frequencies develop 
   during oscillations of different dynamical variables. Whether there is a possibility that a cascade of frequencies 
   develop eventually leading to chaotic behavior remains an open question.
\vskip -2.0 cm
\section{Acknowledgment}
\vskip -0.5 cm
   Useful discussion with Shraddha Mishra is thankfully acknowledged. K.B. acknowledges the support from the BITS 
   Research Initiation Grant Fund.

\end{document}